\documentclass[12pt]{article}
 \pdfoutput=1
\usepackage[utf8]{inputenc}
\usepackage{amsfonts,amsmath}
\usepackage{jheppub}

\usepackage{color}
\usepackage{tabularx}

\usepackage{ifthen}
\usepackage{enumerate}
\usepackage{amsmath}
\usepackage{amssymb}
\usepackage[mathscr]{eucal}
\usepackage[errorshow]{tracefnt}
\usepackage{amsmath}
\usepackage{slashed}
\usepackage{amssymb}
\usepackage{array}
\usepackage{amsthm}
\usepackage{eucal}
\usepackage{epsf}
\usepackage{epsfig}
\usepackage{psfrag}
\usepackage{multirow}
\usepackage{wasysym} 
\usepackage{tikz-cd} 
\usepgfmodule{shapes}
\usetikzlibrary{backgrounds, arrows,calc,shapes,decorations.pathreplacing, decorations.markings, automata,positioning}

\newcommand{\nn}{\nonumber}
\newcommand{\rf}{r_{\phi}}

\newcommand{\beqa}{\begin{eqnarray}}
\newcommand{\eeqa}{\end{eqnarray}}

\def\W{{\cal W}} 
\def\Tr{\rm Tr}

\newcommand{\beq}{\begin{equation}}
\newcommand{\eeq}{\end{equation}}
\newcommand{\bea}{\begin{eqnarray}}
\newcommand{\eea}{\end{eqnarray}}

\newcommand{\CB}{{\mathcal B}}

\newcommand{\CF}{{\mathcal F}}

\newcommand{\CM}{{\mathcal M}}
\newcommand{\M}{{\mathfrak M}}
\newcommand{\CN}{{\mathcal N}}
\newcommand{\CO}{{\mathcal O}}

\newcommand{\CT}{{\mathcal T}}

\newcommand{\CW}{{\mathcal W}}
\newcommand{\CZ}{{\mathcal Z}}

\newcommand\qt{\tilde q}

\newcommand\pt{\tilde p}
\newcommand\Qt{\tilde Q}
\newcommand\Pt{\tilde P}

\newcommand{\be}{\begin{equation}}
\newcommand{\ee}{\end{equation}}

\newcommand{\bpic}{\begin{tikzpicture}}
\newcommand{\epic}{\end{tikzpicture}}

\def\+{{+\!\!\!+}}

\def\a{\alpha} 
\def\b{\beta}

\def\0{\nonumber}

\def\W{{\cal W}}

\def\Tr{{\rm Tr}}

\def\1{{\bf 1}}

\title{Abelianization and Sequential Confinement in $2+1$ dimensions}

\preprint{SISSA 28/2017/MATE-FISI}

\author[1,2]{Sergio Benvenuti}
\author[2,3]{Simone Giacomelli}

\affiliation[1]{International School of Advanced Studies (SISSA), Via Bonomea 265, 34136 Trieste, Italy}
\affiliation[2]{INFN, Sezione di Trieste, Via Valerio 2, 34127 Trieste, Italy}
\affiliation[3]{International Center for Theoretical Physics, Strada Costiera 11, 34151 Trieste, Italy}
\emailAdd{benve79@gmail.com, sgiacome@ictp.it}

\abstract{We consider the lagrangian description of Argyres-Douglas theories of type $A_{2N-1}$, which is a $SU(N)$ gauge theory with an adjoint and one fundamental flavor. An appropriate reformulation allows us to map the moduli space of vacua across the duality, and to dimensionally reduce. Going down to three dimensions, we find that the adjoint SQCD ``abelianizes": in the infrared it is equivalent to a $\mathcal{N}=4$ linear quiver theory. Moreover, we study the mirror dual: using a monopole duality to "sequentially confine" quivers tails with balanced nodes, we show that the mirror RG flow lands on $\mathcal{N}=4$ SQED with $N$ flavors. These results make the supersymmetry enhancement explicit and provide a physical derivation of previous proposals for the three dimensional mirror of AD theories.}

\begin{document}
\maketitle

\newpage
\section{Introduction and summary}

Recently Maruyoshi and Song \cite{Maruyoshi:2016tqk,Maruyoshi:2016aim} discovered 'Lagrangians for Argyres-Douglas theories'. They  coupled $4d$ $\CN=2$ superconformal theories to a chiral field $A$, transforming in the adjoint of the global symmetry group. Giving a nilpotent vacuum expectation value (vev) to $A$ triggers an RG flow. Studying the infrared CFT, they found that sometimes  the RG flow lands on $\CN=2$ Argyres-Douglas theories \cite{Maruyoshi:2016tqk,Maruyoshi:2016aim,Agarwal:2016pjo}.

For instance, in \cite{Maruyoshi:2016aim} it was shown that starting from $SU(N)$ gauge theory with $2N$ flavors, $\CT_{UV}$, a maximal nilpotent vev initiates an RG flow to the $\CN=1$ gauge theory $SU(N)$ with an adjoint and one flavor, plus some gauge-singlet fields. The IR theory, $\CT_{IR}$ ($SU(N)$ with an adjoint and one flavor), turns out to be equivalent in the infrared to the so called $A_{2N-1}$ Argyres-Douglas theory (see \cite{Argyres:1995jj,Argyres:1995xn,Eguchi:1996ds,Eguchi:1996vu} for a detailed discussion about Argyres-Douglas theories) plus a free sector consisting of operators which violate the unitarity bound and decouple \cite{Kutasov:2003iy}.

In this paper we provide two physical mechanisms for this duality going down to $3$ dimensions, generalizing the case of $SU(2)$ dual to $A_3$ studied in \cite{Benvenuti:2017lle}. 

First we need modify the $4d$ Lagrangians, in two ways. As in \cite{Benvenuti:2017lle}, we introduce gauge singlet fields $\b_j$ which implement the decoupling of the operators that violate the $4d$ unitarity bound. This prescription provides a completion of the theory, allows all standard computations, and to preserve the $4d$ duality when going down to $3d$. We would like to stress that this caveat is not related to the phenomenon observed in \cite{Aharony:2013dha}. We will indeed see later that no monopole superpotential is generated in the compactification. More evidence that adding the fields $\b_j$'s is necessary comes from the fact that the $\b_j$'s map to a particular component of the Coulomb branch short multiplets of the $\CN=2$ AD theory. 

As for the second modification, the superpotential written in \cite{Maruyoshi:2016aim} is incorrect, one term must be removed, in order to satisfy a criterion of chiral ring stability \cite{Collins:2016icw,Benvenuti:2017lle}. The standard procedure of keeping all terms consistent with the symmetries in these cases must be improved. 

We call the modified theories $\CT'_{4d,UV}$ and $\CT'_{4d,IR}$. 
We study the dimensional reduction of the RG flow $\CT'_{UV} \rightarrow \CT'_{IR}$, and its mirror dual $\tilde{\CT}_{UV} \rightarrow \tilde{\CT}_{IR}$. Our $3d$ results are summarized in the following diagram\footnote{A circle is a $U(n)$ gauge group, a double circle is a $SU(n)$ gauge group, a square is a flavor group.}:
\begin{small}\bea \label{DIAGRAM0}\nn 
\bpic  \path (-0.8,1.2) node {$\CT'_{UV}$:} -- (-1.5,0) node[circle,draw](x1) {$N$} -- (0,0) node[rectangle,draw](x2) {$2N$} -- (-0.8,-1) node {$\CW = \CW_{\CN=4}+\delta \CW_{\CN=2}(\a_r,\b_j)$}; \node[circle,draw] at (-1.5,0) {$\quad$};
\draw [-] (x1) to (x2);  \epic \qquad  
  & \bpic \draw[<->, thick] (0,0.4) -- (1.4,0.4); \node[right] at (0,0.7){\small{mirror}}; \node[right] at (0.1,-0.5){$\,$};\epic & \qquad   
    \bpic  \path (3,1.2) node {$\tilde{\CT}_{UV}$:} (2,0) node[circle,draw](x1) {\small$1$} -- (3,0) node[circle,draw](x2) {\small$2$} -- (4,0) node(x3) {$\cdots$}-- (5,0) node[circle,draw](x5) {\small$\!N\!$} -- (4.5,1) node[circle,draw,red,thick](x6) {\small$1$} -- (5.5,1) node[rectangle,draw](x7) {$1$}  -- (6,0) node(x8) {$\cdots$} -- (7,0) node[circle,draw](x9) {\small$2$} -- (8,0) node[circle,draw](x10) {\small$1$}-- (5,-1) node {$\CW = \CW_{\CN=4}+\delta \tilde{\CW}_{\CN=2}(\a_r,\b_j)$};
\draw [-] (x1) to (x2);
\draw [-] (x2) to (x3);
\draw [-] (x3) to (x5);
\draw [-] (x5) to (x6);
\draw [-] (x5) to (x7);
\draw [-] (x5) to (x8);
\draw [-] (x8) to (x9);
\draw [-] (x9) to (x10);
  \epic \nn \\ \nn
\qquad\qquad \bpic \draw[->,thick] (0,0.5) -- (0,-1.5); \node[right] at (0.2,0.2) {\small{$\delta\CW_{\CN=2}$ gives}}; \node[right] at (0.2,-0.4) {\small{mass to $2N\!-\!1$ flavors,}}; \node[right] at (0.2,-1) {\small{$\Tr(\qt \phi^{2N}q)$ drops out}};\epic &&\qquad \quad\qquad\qquad 
\bpic \draw[->,thick] (0,0.5) -- (0,-1.5); 
\node[right] at (0.2,0.2) {monopole superpotentials}; 
\node[right] at (0.2,-0.4) {\emph{sequentially confine} the gauge};  
\node[right] at (0.2,-1) {groups in the lower row};
\epic\\
\bpic \path (-3.5,0) node {$\CT'_{IR}:$}   (-2.5,0) node[circle,draw](x1) {\small{$\!N\!$}}  -- (-0.8,0) node[rectangle,draw](x2) {$1$} --
  (-1.5,-1.1) node {\small{$\CW= \sum_{r} \a_r \Tr(\qt \phi^r q)+\sum_j \b_j \Tr(\phi^j)$}}; \node[circle,draw] at (-2.5,0) {$\quad$};
\draw [->] (x1) to[bend left] (x2); \draw [<-] (x1) to[bend right] (x2);
\draw [->] (x1) to[out=-45, in=0] (-2.5,-0.8) to[out=180,in=225] (x1);
\node[circle,draw] at (-2.5,0){$\,\,\,\,$};
\node[below right] at (-2.4,-0.3){$\phi$};
\node[above right] at (-1.7,0.3){$q$};
\node[below right] at (-1.6,-0.3){$\qt$};
\draw[-,thick] (-2.75,-1.5) -- (-2.75,-2.3); \draw[-,thick] (-2.9,-1.5) -- (-2.9,-2.3); \node[right] at(-2.7,-1.9) {$\rf=0$: \emph{Abelianization}};
\path (-4,-2.8) node[rectangle,draw](z1) {\,1\,} -- (-3.2,-2.8) node[circle,draw](z2) {\!1\!} -- (-2.4,-2.8) node(z3) {$\cdots$} -- (-1.6,-2.8) node[circle,draw](z4) {\!1\!} -- (-0.8,-2.8) node[rectangle,draw](z5) {\,1\,} -- (0,-2.8) node[right] {$\CW\!=\! \CW_{\CN=4}$}; 
\draw [-] (z1) to (z2);
\draw [-] (z2) to (z3);
\draw [-] (z3) to (z4);
\draw [-] (z4) to (z5);
\epic
  & \bpic \draw[<->, thick] (0,0.4) -- (1.4,0.4); \node[right] at (0,0.7){\small{mirror}}; \node[right] at (0.1,0.2){$\,$};\epic   &\qquad   \qquad
\bpic  
\node[right] at (0,1.5){$\tilde{\CT}_{IR}$:};
\node[right] at (1.5,1.75) {\small{Model proposed for the}};
\node[right] at (1.5,1.25) {\small{$3d$ mirror of $A_{2N-1}$ AD}};
\node[circle,draw,red,thick](x1) at (0,0){$1$};
\node[rectangle,draw](x2) at (1.5,0){$N$};    
\node[right] at (2.5,0){$\CW = \CW_{\CN=4}$};\node at (2,-0.6) {$\,$};
\draw[-] (x1) -- (x2);
  \epic 
\eea\end{small}
We analyze the left side of this diagram in section \ref{ABELIAN} and the right side in section \ref{SC}.

We exhibit strong evidence that $\CT'_{IR,3d}$ is equivalent in the IR to an $\CN=4$ Abelian $U(1)^{N-1}$ theory. Two different Lagrangian, UV free, theories are dual in the IR. The mechanism of this \emph{Abelianization duality} is that in $3d$ there is an emergent $U(1)$ global symmetry, and the result of $\CZ$-extremization \cite{Kapustin:2009kz,Jafferis:2010un} is that the superconformal r-charge of the adjoint field vanishes: $\rf=0$. Using the input $\rf=0$, we show that the integrand of $\CZ_{S^3}$ reduces to the integrand of the $\CN=4$ $U(1)^{N-1}$ theory.

We present and check numerically a map between the supersymmetric $S^3$ partition functions of the non-Abelian and Abelian theories.
 
We also show that the chiral ring of the $SU(N)$ gauge theory is isomoporphic to the chiral ring of the $\CN=4$ abelian theory, using recent results about dressed monopole operators in $3d$ non-Abelian gauge theories \cite{Cremonesi:2013lqa}. The emergent $U(1)$ symmetry enhances to an $SU(N)$ flavor symmetry and the generators of the dressed monopoles of the $SU(N)$ gauge theory transform in the adjoint representation of the emergent $SU(N)$ flavor symmetry. 

The Abelianization duality we propose is quite peculiar. For instance, in usual dualities, such as Seiberg duality \cite{Seiberg:1994pq} or $3d$ mirror symmetry \cite{Intriligator:1996ex} (see also \cite{Aharony:1997bx}), at least the Cartan generators of the global symmetry group are visible in both descriptions. In our case the Cartans of the emergent $SU(N)$ global symmetry are themselves emergent in the non-Abelian UV description, while in the Abelian theory they are the topological symmetries.

In order to provide further evidence for the claims on the left side of the diagram (\ref{DIAGRAM0}), in section \ref{SC} we study the mirror RG flow $\tilde{\CT}_{UV} \rightarrow \tilde{\CT}_{IR}$, depicted on the right side. In this case we use very recent results for dualities of $3d$ $\CN=2$ $U(N)$ gauge theories with linear monopole superpotentials \cite{Benini:2017dud}. Starting from the $\tilde{\CT}_{UV}$ quiver, the monopole duality implies that all the gauge nodes in the lower row of $\tilde{\CT}_{UV}$ confine one after the other, starting from one $U(1)$ node and ending with the opposite $U(1)$ node. We call this phenomenon \emph{sequential confinement}. It works for quiver tails with balanced nodes starting from an $U(1)$ gauge group, and is the mirror counterpart of integrating out flavors that get mass from the nilpotent vev. The left-over theory in the IR is $\CN=4$ supersymmetric. This makes the supersymmetry enhancement explicit.

In order to illustrate the procedure, we first discuss the 3d mirror of $A_3$ AD theory building on the results found in \cite{Benvenuti:2017lle} and then proceed with the analysis of the general case. The mirror RG flow lands on SQED with $N$ flavors and enhanced  $\CN=4$ supersymmetry; the surviving $U(1)$ is depicted in red. The latter theory is well known to be mirror of the linear quiver $U(1)^{N-1}$, and was proposed to be the mirror of the $3d$ reduction of $A_{2N-1}$ Argyres-Douglas \cite{Nanopoulos:2010bv}, based on mathematical results \cite{Boalch}. The claim of \cite{Nanopoulos:2010bv} passes several nontrivial consistency checks and is perfectly consistent with the structure of the superconformal index \cite{Buican:2015hsa,Buican:2015ina}. Our method clearly explains why the theory abelianizes in 3d.

Using our $3d$ \emph{sequential confinement} interpretation, it is possible to generalize the story, and find a $4d$ Lagrangian for more general Argyres-Douglas models, like the ones arising from $N$ M5's on a sphere with an irregular puncture \cite{BenveGiaco3}. Again, in the $3d$ mirror many nodes sequentially confine, and in the IR the RG flow lands on the Abelian $\CN=4$ theories of  \cite{Nanopoulos:2010bv}.

\subsection{Notation}
\paragraph{Quiver diagrams}
\begin{itemize}
\item a circle node $\bpic \node[circle,draw] at(0,-0.3) {$\tiny{\!\!N\!\!}$}; \epic$ denotes a $U(N)$ gauge group;
\item a double-circle node $\bpic \node[circle,draw] at(0,-0.3) {$\tiny{\!\!N\!\!}$}; \node[circle,draw] at(0,-0.3) {$\quad\,$}; \epic$ denotes a $SU(N)$ gauge group;
\item a square node $\bpic \node[rectangle,draw] at(0,-0.3) {$\,\tiny{N}\,$}; \epic$ denotes a $U(N)$ or $SU(N)$ flavor group;
\item sometimes we use an $8$-supercharges notation {\tiny $\bpic \node[circle,draw](x) at(0,-0.3) {$\tiny{\!\!N_1\!\!}$}; \node[circle,draw](y) at(0.9,-0.3) {$\tiny{\!\!N_2\!\!}$}; \draw[-] (x) to (y); \epic$}, links are bifundamental hypers and adjoints in the vector multiplets are implicit;
\item sometimes we use a $4$-supercharges notation {\tiny$\bpic \node[circle,draw](x) at(0,-0.3) {$\tiny{\!\!N_1\!\!}$}; \node[circle,draw](y) at(1.1,-0.3) {$\tiny{\!\!N_2\!\!}$}; \draw [->] (x) to[bend left] (y); \draw [<-] (x) to[bend right] (y); \draw [->] (x) to[out=-45, in=0] (0,-0.8) to[out=180,in=225] (x);\epic$}, arrows are bifundamental or adjoint chiral fields.\end{itemize}

\paragraph{Flips}
A gauge singlet chiral field $\sigma$ \emph{flips an operator} $\CO$ when it enters the superpotential through the term $\sigma\cdot \CO$. In this paper we consistently use different names for three classes of flipping fields:
\begin{itemize}
\item $\a_r$ fields flip the dressed mesons operators, which are mapped to monopole operators $\M$ with topological charges $(0,\ldots,0,-,\ldots,-,0,\ldots,0)$ in the mirror quiver.
\item $\b_j$ fields flip $\Tr(\phi^j)$, which are mapped to length-$j$ mesons in the mirror quiver.
\item $\gamma_N$ fields are generated  in the mirror quiver when gauge nodes confine. They flip the $N \times N$ determinant of the dual Seiberg mesons.
\end{itemize}

\section{Adjoint-SQCD with one flavor in $3d$: \emph{Abelianization}}\label{ABELIAN}

The starting point is $4d$ $\CN=2$ $SU(N)$ gauge theory, with $2N$ flavors $q^i,\qt_i$ and an additional singlet field $A$ in the adjoint of the global symmetry $SU(2N)_F$, coupled to the moment map $\mu_H=\Tr( \qt_i q^j )$. Notice that the latter coupling is marginally irrelevant and explicitly breaks $\CN=2$ supersymmetry to $\CN=1$. \cite{Maruyoshi:2016tqk,Maruyoshi:2016aim} then gave a maximal $2N \times 2N$ nilpotent vev to $A$. We review the procedure of integrating out the massive flavors due to the nilpotent vev \cite{Gadde:2013fma,Agarwal:2014rua} in Appendix \ref{nilpotvev}. The nilpotent vev breaks the $SU(2N)$ flavor symmetry completely and leads to a $\CN=1$ $SU(N)$ gauge theory with an adjoint field $\phi$ and one flavor $q,\qt$:

\bea \label{flowbasic} \bpic  \path (-2,0) node {$\CT_{4d, UV}:$} (0,0) node[circle,draw](x1) {$N$} -- (1.5,0) node[rectangle,draw] (x2) {$2N$} -- (6,0) node {$\CW=\CW_{\CN=2} + \sum_{i,j=1}^{2N} A_{\, j}^{i} \Tr( \qt_i q^j )$};\node[circle,draw] at (0,0) {$\quad$};
\draw [-] (x1) to (x2); \nn
  \epic \qquad \\
     \bpic \draw[->, thick] (0,0.5) -- (0,-1.5);  \node[right] at (0.3,-0.5){maximal nilpotent vev to $A$};  \epic \qquad \qquad \qquad   \\
 \bpic  \path (-4,0) node {$\CT_{4d, IR}:$}   (-2.5,0) node[circle,draw](x1) {$N$}  -- (-0.5,0) node[rectangle,draw](x2) {$\,1\,$} --
  (4.5,0.3) node {$\CW_{IR}=  \Tr(\qt\phi^{2N}q) + \sum_{r=0}^{N-2} \a_r \Tr(\qt \phi^r q)$} --
  (4.6,-0.4) node {$\Tr(\phi^j), \, j=2,3,\ldots N ,$ are decoupled}; \node[circle,draw] at (-2.5,0) {$\quad$};
\draw [->] (x1) to[bend left] (x2); \draw [<-] (x1) to[bend right] (x2);
\draw [->] (x1) to[out=-45, in=0] (-2.5,-1) to[out=180,in=225] (x1);
\node[below right] at (-2.4,-0.6){$\phi$};
\node[above right] at (-1.5,0.3){$q$};
\node[below right] at (-1.4,-0.3){$\qt$};
  \epic \nn \eea

In $\CT_{4d,IR}$ the field $\phi$ has R-charge $R[\phi]=\frac{2}{3(N+1)}$, as determined applying A-maximization. One important aspect of A-maximization is that the $N-1$ gauge invariant operators $\Tr(\phi^j)$ with $j=2,3,\ldots,N$ have $R<\frac{2}{3}$ and must be decoupled. The $N-1$ singlet fields $\a_r$ are what is left-over from the $2N \times 2N$ matrix $A$. 

Because of the singlets and the peculiar superpotential, the qualitative behavior of the $\CT_{IR}$ is quite different from the case of adjoint-SQCD with $\CW=\Tr(\phi^h)$ studied in the literature \cite{Kutasov:1995np,Kutasov:1995ss,Kim:2013cma,Nii:2014jsa,Amariti:2014iza,Amariti:2016kat}, both in $4d$ and in $3d$. 

An important consequence of the decoupling of all the operators $\Tr(\phi^j)$ with $j=2,3,\ldots,N$ is that $\phi^N$, as $N \times N$ matrix, is zero in the chiral ring. This implies a truncation in the spectrum of  gauge invariant operators like mesons and baryons dressed by the adjoint fields. 

In particular, dressed mesons $\Tr(\qt\phi^{r}q)$ vanish if $r \geq N$, so the first term in the superpotential $\Tr(\qt\phi^{2N}q)$ is zero in the chiral ring. This in turn implies that the chiral ring as defined by the lower theory in (\ref{flowbasic}) is \emph{unstable}.

Let us state in detail the criterion of chiral ring stability as in \cite{Benvenuti:2017lle}. Starting from a theory $\CT$ with superpotential $\CW_\CT = \sum_i \CW_i$ (where each term $\CW_i$ is gauge invariant), one needs, for each $i$, to:
\begin{itemize}
\item consider the modified theory $\CT_i$, where the term $\CW_i$ is removed from $\CW$
\item check if the operator $\CW_i$ is in the chiral ring of $\CT_i$
\end{itemize}
If one of the terms $\CW_i$ does not pass the test, it must be discarded from the full superpotential $\CW_\CT$.
See \cite{Benvenuti:2017lle} for a more detailed justification of this procedure and \cite{Collins:2016icw} for a geometric interpretation in terms of K-stability. 

If we drop $\Tr(\qt\phi^{2N}q)$ from the superpotential, then $\Tr(\qt\phi^{2N}q)$ is still zero in the modified chiral ring, so $\Tr(\qt\phi^{2N}q)$ does not pass the test of chiral ring stability: the correct IR superpotential does not contain the term $\Tr(\qt\phi^{2N}q)$.\footnote{Notice that if one believes that the term $\Tr(\qt\phi^{2N}q)$ can appear in the superpotential, then there would be an exactly marginal direction (this is because $\Tr(\qt\phi^{2N}q)$ has R-charge $2$ and does not break any non-anomalous global symmetry, for $N>2$, so it generates an exactly marginal direction \cite{Kol:2002zt, Benvenuti:2005wi,Green:2010da,Kol:2010ub}), but in the $A_{2N-1}$ AD model there are no marginal directions. For $N=2$, the term $\Tr(\qt\phi^{4}q)$ breaks the global $SU(2)$ symmetry which must be present in the $A_3$ AD model \cite{Benvenuti:2017lle}.} 

Moreover, in order to reduce to $3d$, it is crucial that we do not simply reduce the $\CN=2$ $SU(N)$ with $2N$ flavors theory and then repeat the same procedure in $3$ dimensions \cite{Benvenuti:2017lle}: this strategy would lead to a different set of flipping $\a_r$ fields coupled to the $3d$ IR theory. For instance in the case of $SU(N=2)$, in \cite{Benvenuti:2017lle} it was shown that, repeating the procedure of giving a maximal nilpotent vev to $A$ in $3d$, the IR theory contains also a flipping term $\a_1 \Tr(\qt \phi q)$, and instead of being dual to $\CN=4$ $U(1)$ with $2$ flavors, the IR theory contains two gauge singlets and is dual to $\CN=2$ $U(1)$ with $2$ flavors with both flavors flipped. For general $N$, in Appendix \ref{nilpotvev}, using the chiral ring stability criterion, we show that at most $N$ $\a_r$ gauge-singlets ($r=0,1,\ldots,N-1$) from the $2N \times 2N$ matrix $A$ can stay attached to the theory. In $4d$ $a$-maximization imposes that $\a_{N-1}$ decouples, while in $3d$ we have a choice of keeping $\a_{N-1}$ in the theory or not. If we perform the Maruyoshi-Song procedure in $3d$, all $N-1$ $\a_r$'s remain in the IR, and as we discuss in more detail in section \ref{MS3d}, the low energy theory is not $\CN=4$ supersymmetric.

We thus introduce in the UV precisely $N-1$ $\a_r$ gauge singlets fields and also the $N-1$ $\b_j$ fields to flip the operators $\Tr(\phi^j)$. In this way the UV description is complete and in the IR there in no unitarity violation.

We call the modified theories $\CT'_{4d,UV}$ and $\CT'_{4d,IR}$, and replace \ref{flowbasic} with:

\bea \label{flowT'} \bpic   \path (-4,0) node {$\CT'_{4d, UV}:$}   (-2.5,0) node[circle,draw](x1) {$N$}  -- (-1,0) node[rectangle,draw](x2) {$2N$}
 -- (1,0.4) node[right] {$\CW_{UV}=  \sum_{i=1}^{2N} \Tr(\qt_i \phi q^i) + \sum_{i=1}^{2N-1} \Tr(\qt_i q^{i+1})+$}
 -- (1.7,-0.4) node[right] {$ + \sum_{r=0}^{N-2}\sum_{i=0}^r \a_r \Tr(\qt_{2N+i-r}q^{i+1}) + \sum_{j=2}^{N} \b_j \Tr(\phi^j)$};
\node[circle,draw] at (-2.5,0) {$\quad$};
\draw [->] (x1) to[bend left] (x2); \draw [<-] (x1) to[bend right] (x2);
\draw [->] (x1) to[out=-45, in=0] (-2.5,-1) to[out=180,in=225] (x1);
\node[below right] at (-2.4,-0.6){$\phi$};
\node[above right] at (-2,0.3){$q_i$};
\node[below right] at (-1.9,-0.3){$\qt^i$};
  \epic \nn \\
    \bpic \draw[->, thick] (0,0.5) -- (0,-1.5);  \node[right] at (0.3,-0.2){Integrate out the $2N-1$ massive};  \node[right] at (0.3,-0.8){flavors $q_1,q_2,\ldots,q_{2N-1}$, $\qt_2,\qt_3,\ldots,\qt_{2N}$}; \epic \qquad  \qquad   \\
 \bpic  \path (-3.7,0) node {$\CT'_{4d, IR}:$}   (-2.5,0) node[circle,draw](x1) {$N$}  -- (-1,0) node[rectangle,draw](x2) {$\,1\,$} --
  (4.5,0) node {$\CW_{IR}=  \sum_{r=0}^{N-2} \a_r \Tr(\qt \phi^r q) +  \sum_{j=2}^{N} \b_j \Tr(\phi^j)$};
  \node[circle,draw] at (-2.5,0) {$\quad$};
\draw [->] (x1) to[bend left] (x2); \draw [<-] (x1) to[bend right] (x2);
\draw [->] (x1) to[out=-45, in=0] (-2.5,-1) to[out=180,in=225] (x1);
\node[below right] at (-2.4,-0.6){$\phi$};
\node[above right] at (-2,0.3){$q$};
\node[below right] at (-1.9,-0.3){$\qt$};
  \epic \nn \eea

In $\CT'_{4d}$ only the flipping fields $\a_r$ with $r=0,1,\ldots,N-2$ are present in the UV definition of the theory, and we also introduced $N-1$ flipping fields $\b_j$, that survive in the IR. As discussed in more detail in \cite{Benvenuti:2017lle}, this operation has precisely the same effect of stating that the operators $\Tr(\phi^j)$ are decoupled as in \cite{Kutasov:2003iy}. The $3d$ and $4d$ superconformal indices, the $S^3$ partition function and $4d$ $a$-maximization \cite{Intriligator:2003jj} are all the same. One advantage of this "completed" re-formulation is that now standard techniques can be used to compute the chiral ring and the moduli space of vacua. $\CT'_{4d}$ can also be easily compactified on a circle.

\subsection{$4d$ chiral ring: dressed baryons and dressed meson}\label{dressed1}
Before compactifying to $3d$, let us study the chiral ring of the $4d$ theory. The theory admits two non-anomalous global symmetries, acting on the elementary fields as
\be\label{char21}
\begin{array}{c|ccc}
 &  U(1)_{R}^{4d} &  U(1)_T  & U(1)_{B}  \\ \hline
\phi & \frac{2}{3(N+1)} &  \frac{2}{3(N+1)}  & 0  \\
q,\qt & \frac{1}{3}+\frac{2}{3(N+1)} &  -\frac{2N}{3(N+1)} &  \pm 1 \\ 
\b_j &  2-\frac{2j}{3(N+1)} & - \frac{2j}{3(N+1)} &   0 \\ 
\a_r &  \frac{4N-2r}{3(N+1)} & \frac{4N-2r}{3(N+1)} &   0 \\ 
\end{array}
\ee
where we normalized the non baryonic global symmetry $U(1)_T$ so that $R[\phi]=T[\phi]$ and $R[\a_r]=T[\a_r]$. Notice that $R[\b_{r+2}]=\frac{6N+2-2r}{3(N+1)}=R[\a_r]+\frac{2}{3}$.

As pointed out in \cite{Maruyoshi:2016aim}, the $N-1$ $\a_r$'s, $r=0,1,\ldots,N-2$, map to the Coulomb branch generators of $A_{2N-1}$ AD. Let us study the rest of the chiral ring. 

First of all we claim that the operators $\b_j$ vanish in the chiral ring: they are Q-exact operators, where Q denotes the supercharges which emerge in the infrared. We postpone the discussion about this point to the end of this subsection; for the moment we just point out, as a consistency check, that they cannot have an expectation value: such a vev would lead to a theory with no vacuum for quantum reasons.
 
For instance, if $\b_2$ takes a vev, $\phi$ becomes massive, and the low energy theory is $\CN=1$ $SU(N)$ with $1$ flavor and $\CW=\a_0 \Tr(\qt q)$, which has no vacuum because a ADS superpotential is dynamically generated.
 
For a generic $j \leq N$, giving vev to $\b_j$ brings us to a theory with $\CW=\Tr(\phi^j)$. \cite{Kutasov:1995np} showed that a $SU(N)$ gauge theory with $N_f$ flavors and $\CW=\Tr(\phi^j)$ has a vacuum only if $N_f \geq \frac{N}{j-1}$. Since we have $N_f=1$, giving a vev to $\b_j$ leads to a theory with no vacuum, for all $j=2,3,\ldots,N$.

Assuming that all $\b_j$'s vanish in the chiral ring, and using the powerful matrix relation $\phi^N=0$, it is quite easy to discuss the full structure of the $4d$ chiral ring.

The operators that are built using $q,\qt$ and $\phi$ are generated by only three operators. Since $\phi^N=0$, we can make only one \emph{dressed baryon}, using $N$ $q$ fields and $\binom{N}{2}$ $\phi$ fields as follows
  \be \CB=\varepsilon_{i_1,i_2, \ldots, i_{N}}\,\, q^{i_1} \, (\phi q)^{i_2} \, (\phi^2 q)^{i_3} \, \ldots \, (\phi^{N-1}q)^{i_{N}} \ee
  with
  \be R[\CB]= -2T[\CB] =\frac{2}{3}N\ee
There is a similarly defined anti-baryon $\tilde \CB$ using $\qt$. See \cite{Hanany:2008sb} for the Hilbert Series of adjoint SQCD with $N_f$ flavors. Because of the $\CF$-terms of $\a_r$ and the relation $\phi^N=0$, there is only one non-vanishing dressed meson:
\be \CM=\Tr(\qt \phi^{N-1} q)\ee
with
\be R[\CM]=-2T[\CM]=\frac{4}{3}\ee
$\CB, \tilde \CB$ and $\CM$ satisfy the chiral ring relation
 \be\label{ccad} \CB \cdot \tilde{\CB} = \varepsilon_{i_1,i_2, \ldots, i_{N}}\,\varepsilon^{j_1,j_2, \ldots, j_{N}}\, q^{i_1} \, (\phi q)^{i_2} \, \ldots \, (\phi^{N-1}q)^{i_{N}}\, \qt_{j_1} \, (\qt \phi)_{j_2} \, \ldots \, (\qt \phi^{N-1})_{j_{N}}=\CM^{N} \,,\ee
 where we used that $\Tr(\qt \phi^{r} q)=0$ in the chiral ring if $r<N-1$.

The chiral ring relation $\CB \cdot \tilde{\CB} = \CM^N$ is precisely the defining equation of $\mathbb{C}^2/\mathbb{Z}_N$, known to be the Higgs branch of $A_{2N-1}$ Argyres-Douglas.

The other generators of the chiral ring are the $N-1$ gauge singlets $\a_r$, and map to the Coulomb branch of $A_{2N-1}$ Argyres-Douglas. Let us study the chiral ring relations between the $\a_r$'s and $\CB,\CM,\tilde{\CB}$. Contracting the $\CF$-terms of $\qt_i$
\be \label{qtFterm}\sum_s \a_s (\phi^s q)^i = 0 \ee
with $(\qt \phi^{N-1-r})_i$ we find
\be \a_r \cdot \CM= 0\qquad \text{for every}\,\, r\ee
Contracting (\ref{qtFterm}) with 
$$\varepsilon_{i_0,\ldots,i_{r-1},i, i_{r+1},\ldots,i_{N-1}}\,\, q^{i_0} (\phi q)^{i_1} \ldots (\phi^{r-1}q)^{i_{r-1}} (\phi^{r+1}q)^{i_{r+1}} \ldots (\phi^{N-1}q)^{i_{N-1}} $$
we find
\be \a_r \cdot \CB =0\qquad \text{for every}\,\, r\ee
Similarly one can prove that
\be  \a_r \cdot \tilde{\CB}= 0\qquad \text{for every}\,\, r\ee
Concluding the $\a_r$'s have vanishing product with the three generators $\CB,\CM,\tilde{\CB}$. There are no relations involving only the $\a_r$'s.

So the $4d$ moduli space of vacua has two branches: one branch is $\mathbb{C}^{N-1}$, freely generated by the $N-1$ $\a_r$'s, the other branch is $\mathbb{C}^2/\mathbb{Z}_N$. The two branches intersect only at the origin of the moduli space. This is precisely the expected moduli space of vacua of the $A_{2N-1}$ Argyres-Douglas theory. 

\subsubsection{$\CN=2$ AD interpretation of the $\beta_j$ multiplets} 

The R-symmetry of any $\CN=2$ SCFT is $SU(2)_R \times U(1)_{R_{\CN=2}}$. The R-symmetry of an $\CN=1$ subalgebra is given by the combination 
\be R_{\CN=1}=\frac{1}{3}R_{\CN=2}+\frac{4}{3}I_3 \ee
where $I_3$ is the cartan generator of $SU(2)_R$. The supercharges $Q_{\alpha}$ generating this $\CN=1$ subalgebra (together with the corresponding $\bar{Q}_{\dot{\alpha}}$) are those with charge $\frac{1}{2}$ under $I_3$. In this way the scaling dimension of the $\CN=1$ chiral primaries (defined w.r.t. the above mentioned $Q_{\alpha}$ supercharges) satisfy $\Delta=\frac{3}{2}R_{\CN=1}$. Instead, the only combination under which the gluinos in a Lagrangian $\CN=2$ SCFT are uncharged is proportional to 
\be R_{\CN=2}-2I_3 \ee 
The above mentioned $Q_{\alpha}$ and $\bar{Q}_{\dot{\alpha}}$ supercharges are the only manifest supercharges in the lagrangian description of Argyres-Douglas theories.

The AD theory of type $A_{2N-1}$ contains Coulomb Branch operators, usually called $u_k$, of dimension \be \Delta(u_k)=1+\frac{k}{N+1},\quad k=1,\dots,N-1.\ee 
The $u_k$ operators transform in the trivial representation of $SU(2)_R$, so they have charges
\be R_{\CN=2}[u_k]= 2+\frac{2k}{N+1} \,,\qquad I_3[u_k]=0\ee 
Since AD theories have $\CN=2$ supersymmetry, the $u_k$ operators are the lowest components of short $\CN=2$ supermultiplets, which, in the Dolan-Osborn notation \cite{Dolan:2002zh} are called $\mathcal{E}_{(R_{\CN=2},0,0)}$. We denote the corresponding $\CN=2$ multiplets as $U_k$. As we have already explained, the $u_k$ map to the lowest components of the chiral multiplets $\alpha_{N-1-k}$ in the nonabelian $SU(N)$ theory:
\be \a_{N-1-k} \longleftrightarrow {\bf u}_k\ee
 where ${\bf u}_k$ denotes the $\CN=1$ chiral multiplet one gets acting with the supercharge $Q_{\alpha}$ on the chiral primary $u_k$. The chiral multiplets $\alpha_{N-1-k}$ represent only half of the $U_k$ CB multiplets and the remaining components are obtained by acting with the ``hidden'' supercharges, which have charge $-\frac{1}{2}$ under $I_3$ (see also \cite{Giacomelli:2014rna} for a discussion about this point). These extra components are organized into another $\CN=1$ chiral multiplet (where again chirality refers to the $Q_{\alpha}$ supercharges described before) which we call ${\bf v}_k$: 
 \be {\bf v}_k\equiv\int d^2\tilde{\theta}U_k\,,\ee
where  $\tilde{\theta}$ represent the IR emergent Grassmann variables of the $\CN=2$ superspace (the notation is identical to that of \cite{Argyres:1995xn}). The $\tilde{\theta}$'s have charge 1 under $R_{\CN=2}$ and -1/2 under $I_3$, so
\be R_{\CN=2}[{\bf v}_k]= \frac{2k}{N+1} \,,\qquad I_3[{\bf v}_k]=1\ee 
The charge under the R-symmetry of the manifest $\CN=1$ subalgebra is then
\be R_{\CN=1}[{\bf v}_k] = \frac{4N+4+2k}{3N+3},\quad k=1,\dots,N-1\ee 
This fits perfectly with the R-charge of the $\beta_j$ fields given in (\ref{char21}), once we set $j=N+1-k$. Also the charges of the various fields under $U(1)_T$ in (\ref{char21}), which can be identified with the combination $R_{\CN=2}/3-2I_3/3$ in the $\CN=2$ theory, and $U(1)_B$ are consistent with the claim that $\beta_j$ and $\alpha_{j-2}$ are part of the same $\CN=2$ multiplet. We therefore propose the complete identification 
\bea \a_{r} & \quad \longleftrightarrow \quad& {\bf u}_{N-1-r} \nn\\
&&\bpic  \draw[<->,thick] (0,0) -- (0,-1); \node[right] at (0.15,-0.5) {$\CN=2$ supercharges}; \node at (0,0.2) {};\epic\\ 
 \b_{r+2} & \longleftrightarrow & {\bf v}_{N-1-r} \nn \eea
In other words, $\b_{r+2}$ is a supersymmetric partner of $\a_r$, for an emergent supersymmetry.

The triviality in the chiral ring of $\beta_j$'s now simply follows from the fact that in the $\CN=2$ AD model they are Q-exact.

\subsection{Compactification to $3d$: emergent symmetry}\label{compact}
We now compactify on $S^1$ the RG flow \ref{flowT'}.

First of all, can monopole superpotential be generated? Since the theory contains an adjoint field $\phi$, in order to possibly generate a monopole superpotential, two zero modes must be soaked up by the $4d$ superpotential \cite{Nii:2014jsa}. Terms proportional to $\a_r$ cannot be generated because in a $SU(N)$ theory monopole operators $\M_{SU(N)}$ cannot be dressed with fundamental fields $q,\qt$ and because all dressed mesons $\Tr(\qt \phi^r q)$ vanish in the chiral ring if $r<N-1$. Terms proportional to $\b_j$ cannot be generated because $\b_j=0$ in the chiral ring\footnote{Here we are assuming that the $4d$ result $\b_j=0$ holds also in $3d$, it would be nice to prove this statement.}, so terms like $\b_j\{\M_{SU(N)}\phi^{j-2}\}$ (we denote by $\{\M_{SU(N)}\phi^{i}\}$ the monopole operators dressed by $i$ factors of the adjoint field) would lead to an unstable chiral ring. We conclude that no monopole superpotential is generated in the compactification.

So the $3d$ IR superpotential is the same as in $4d$:
\be\CW_{IR}=  \sum_{r=0}^{N-2} \a_r \Tr(\qt \phi^r q) +  \sum_{j=2}^{N} \b_j \Tr(\phi^j)\ee
This fact has the important consequence that in $3d$ there is an emergent symmetry, on top of the $4d$ symmetries.
\be\label{char3d}
\begin{array}{c|cccc}
 &  U(1)_{R} &  U(1)_q  & U(1)_{T'}  & U(1)_{B}  \\ \hline
\phi & \rf & 0 & \frac{1}{N-1} & 0  \\
q,\qt & r_q &  \frac{1}{2} & -\frac{1}{2} &  \pm \frac{1}{N} \\ 
\b_j &  2-j\!\cdot\!\rf & 0 & -\frac{j}{N-1} &   0 \\ 
\a_r &  2\!-\!2r_q\!-\!r\!\cdot\!\rf & -1 &  1-\frac{r}{N-1} &  0 \\ 
\M_{SU(N)} &  2-2r_q-2(N\!-\!1)\rf & -1 & -1 &   0 \\ 
\end{array}
\ee
$T'$ is chosen so that the baryons $\CB, \tilde{\CB}$ and the meson $\CM$ are neutral. The basic monopole operator $\M_{SU(N)}$ has GNO charges $\{+1,0,\ldots,0,-1\}$. In any $3d, \CN=2$ $SU(N)$ gauge theory with an adjoint field $\phi$, a fundamental $q$ and an anti-fundamental $\qt$, the monopole global symmetry charges can be computed in terms of the charges of the elementary fermionic fields in the lagrangian
\be F[\M_{SU(N)}] = - F[q] - F[\qt] - 2(N-1)F[\phi] \ee
\be R[\M_{SU(N)}] = 1-R[q]+1-R[\qt]+2(N-1)(1-R[\phi])-2(N-1) \ee

\subsection{$\CZ$-extremization: Abelianization}
Let us study the $S^3$ partition function. The contribution of chiral field with r-charge $r$ is $e^{l(1-r)}$. The function $l(x)$ is defined as follows: 
\be l(x)=-xlog\left(1-e^{2\pi ix}\right)+\frac{i}{2}\left(\pi x^2+\frac{1}{\pi}\text{Li}_2(e^{2\pi i x})\right)-\frac{i\pi}{12}\ee
and satisfies the differential equation $\partial_x l(x) =- \pi x cot (\pi x)$. The $S^3$ partition function for  $SU(N)$ with an adjoint of r-charge $r_{\phi}$ and a flavor $q,\qt$ of r-charge $r_q$ is
\bea \CZ_{SU(N)}[r_{\phi},r_q,b]&=&\nn
 \prod_{r=0}^{N-2}e^{l(1-(2-2r_q-r \cdot \rf))} \prod_{j=2}^Ne^{l(1-(2-j \cdot r_{\phi}))}  \int_{-\infty}^{+\infty}\!\!\frac{\prod_{i>j} (2 sinh(\pi(z_i-z_j)))^2 }{N!} \cdot \\
&& \,\,\cdot \, e^{(N-1) l(1-\rf)} \prod_{i\neq j} e^{l(1-r_{\phi} + i (z_i-z_j))}\prod_i e^{l(1-r_q \pm b - i z_i)} \delta(\sum z_i)dz_i \eea
In the first line there is the contribution of the singlets $\a_r$ and $\b_j$, the Haar measure and the $N!$ Weil-group factor. In the second line the contribution of the adjoint field $\phi$ and the fundamental fields $\qt,q$ appear. $b$ is the fugacity for the baryonic symmetry.

Performing $\CZ$-extremization, we find that $\CZ_{SU(N)}$ has a critical point at
\be \rf=0\,,\qquad r_q=\frac{1}{2} \ee
We checked this claim numerically for $N=2,3$. Since the baryonic symmetry doesn't mix with the R-symmetry, the critical point is obviously at $b=0$.

The following limit\footnote{This can be proven as follows: using the explicit expression for $l(z)$ and the identity $l(z)+l(-z)=0$ one can easily derive the equation $l(1+ix)+l(1-ix)=-2log ( 2sinh( \pi x)),$ which immediately implies the desired result.}
\be lim_{\rf \rightarrow 0} e^{l(1-r_{\phi} \pm i x)} (2 sinh( \pi x))^2 = 1\ee 
implies that in the limit $\rf\rightarrow 0$ the off-diagonal components of the adjoint $\phi$ cancel against the Haar measure. The limit\footnote{From the equation $\partial_x l(x) =- \pi x cot (\pi x)$ we get $\partial_x l(1-x) =\pi(x-1) cot (\pi x)$, which implies the asymptotics $l(1-x)\sim-log (sin(\pi x))$ around $x=0$. Using this result and the identity $l(x)+l(-x)=0$, we conclude that 
$$e^{l(1 - (2 - j \rf)) + l(1 - \rf)}\sim e^{log (sin(j\pi\rf))}/e^{log (sin(\pi\rf))}$$ and the r.h.s. manifestly tends to $j$ for $\rf \rightarrow 0$.}
\be lim_{\rf \rightarrow 0}e^{l(1 - (2 - j \rf)) + l(1 - \rf))}=j  \ee
instead implies that the $N-1$ diagonal components of $\phi$ combine with the $N-1$ $\b_j$ fields to cancel the $N!$ Weil-group factor. 

In the limit $\rf \rightarrow 0$ the integrand of the partition function for $SU(N)$ becomes the integrand of the partition function for an Abelian $U(1)^{N-1}$ gauge theory.

The Abelian gauge theory is the $\CN=4$ supersymmetric linear quiver with $N-1$ gauge groups and $3N$ chiral fields $\Phi_i, P_i, \Pt_i$
\be \label{N=4quiver}\bpic \path (-4,-3.2) node[rectangle,draw](z1) {\,1\,} -- (-3,-3.2) node[circle,draw](z2) {\!1\!} -- (-2,-3.2) node(z3) {$\cdots$} -- (-1,-3.2) node[circle,draw](z4) {\!1\!} -- (-0,-3.2) node[rectangle,draw](z5) {\,1\,} -- (2,-3.2) node[right] {$\CW\!=\! \CW_{\CN=4}=\sum_{i=1}^{N-1} \Phi_i (P_i\Pt_i - P_{i+1}\Pt_{i+1} ) $}; 
\draw [-] (z1) to (z2);
\draw [-] (z2) to (z3);
\draw [-] (z3) to (z4);
\draw [-] (z4) to (z5);
\epic\ee
whose most general partition function depends on $r_P$, a baryonic-like fugacity $B$ and $N\!-\!1$  Fayet-Iliopoulos parameters $\eta_j$
\be \CZ_{U(1)^{N-1}}[r_P,B,\eta_i] = e^{(N-1) l(1-(2-2r_P))} \int_{-\infty}^{+\infty}  \prod_{i=1}^N e^{l(1-r_P \pm B - i (z_i\!-\!z_{i+1}))} e^{2 \pi \sum \eta_j z_j}\prod_{j=1}^{N-1} dz_j \ee

Notice that the reduction is at the level of the integrands, which is somehow stronger than the equality at the level of the integrals. The reduction holds on the two-dimensional locus $\rf=\eta_j=0$.

\subsection{$3d$ chiral ring: dressed monopoles}\label{dressed2}
In order to gain better understanding of the Abelianization, we study the complete $3d$ chiral ring, and map it to the chiral ring of the Abelian theory.

Compared to $4d$, in $3d$ there are also monopole and dressed monopole operators. 

$\M_{SU(N)}$ can be dressed with the adjoint field $\phi$. Dressed monopoles were studied in order to compute the Coulomb branch of $\CN=4$ gauge theories in \cite{Cremonesi:2013lqa}, using Hilbert Series techniques \cite{Benvenuti:2006qr}. Formula $(5.11)$ of \cite{Cremonesi:2013lqa} gives the 'Plethystic Logarithm' of the Coulomb Branch Hilbert Series for $\CN=4$ $SU(N_c)$ gauge theories with $N_f$ flavors:
\be \label{PLog}PLog[H_{SU(N_c),N_f}(t)]= \sum_{j=1}^{N_c-1}\left[t^{j+1}+j(t^{N_f-j}+t^{N_f-2N_c+j+1})\right]-t^{2N_f-4N_c+6}+\CO(t^{2N_f-4N_c+7})\ee
The first term represents the algebraic (linearly independent) generators of the Coulomb Branch chiral ring:
\begin{itemize}
\item $\sum_{j=1}^{N_c-1}t^{j+1}$ represents $N_c-1$ generators, with scaling dimension $2,3,\ldots,N_c-1$. These are the $\Tr(\phi^j)$.
\item $\sum_{j=1}^{N_c-1}j(t^{N_f-j}+t^{N_f-2N_c+j+1})$ represents $N_c (N_c-1)$ generators. These are \emph{dressed monopoles} made out of the basic monopole $\M_{SU(N_c),\, \CN=4}$ and $k$ factors of the adjoint field $\phi$. We denote such operators $\{\M_{SU(N_c)}\phi^k\}$. The basic monopole $\M_{SU(N_c)}$ has GNO charges $(+1,0,\ldots,-1)$ and in the $\CN=4$ theory it has scaling dimension $\Delta[\M_{SU(N_c),\,\CN=4}]=N_f-2N_c+2$. Rewriting the sum as $\sum_{j=1}^{N_c-1}j(t^{\Delta_\M+2N_c-2-j}+t^{\Delta_\M+j-1})$, we see that the allowed values for $k$ are $$0, 1, 1, 2, 2, 2, 3, 3, 3, 3, \ldots, 2N_c-4,2N_c-4, 2N_c-3\,.$$
\end{itemize}
The second term in (\ref{PLog}), with a minus sign, represents an algebraic non-linear relation satisfied by the generators, but it is valid only for $\CN=4$ theories, in our $\CN=2$ case it does not apply.

The counting of the generators instead applies to our $\CN=2$ case as well, even though the above results were derived for $\CN=4$ gauge theories: there are $N(N-1)$ linearly independent dressed monopoles\footnote{Giving a vev to $\M_{SU(N)}$ breaks the gauge symmetry $SU(N)\rightarrow U(1)\times SU(N-2) \times U(1)$. The adjoint field $\phi$ decomposes as $\text{diag}(\phi_1,\hat\phi,\phi_N)$, where $\phi_1$ and $\phi_N$ are scalars and $\hat\phi$ is a traceless $N\!-\!2 \times N\!-\!2$ matrix. How many independent ways are there to dress $\M_{SU(N)}$ with $j$ factors of $\phi$'s? We need to consider operators of the form
\be \{ \M_{SU(N)} \phi_1^a \hat\phi^b \phi_N^c \} \, \qquad a+b+c=j \label{dressedmonform}\ee
We need to impose that they cannot be written as a product of $\Tr(\phi^d)$ times some smaller dressed monopole $\{\M_{SU(N)}\phi^{d'}\}$ and we need also to consider the constraints that $\Tr(\phi^{j\geq N})$ can be expressed as a combination of $\Tr(\phi^{i < N})$. Doing this type of analysis, one concludes that there are precisely $N(N-1)$ dressed monopoles generators, see section 5 of \cite{Cremonesi:2013lqa}.}.

In our case of $SU(N)$ theory with one flavor and superpotential 
$$\CW_{IR}=  \sum_{r=0}^{N-2} \a_r \Tr(\qt \phi^r q) +  \sum_{j=2}^{N} \b_j \Tr(\phi^j)$$
the $N-1$ Coulomb branch generators $\Tr(\phi^j)$ are removed by the $\CF$-terms of $\b_j$, but we can combine the dressed monopoles with the $N-1$ $\a_r$'s. All together there are $N^2-1$ operators with $U(1)_q$-charge $-1$ and vanishing baryonic charge $U(1)_B$. Using the input $\rf=0$, $r_q=\frac{1}{2}$, the scaling dimension of all these $N^2-1$ operators is $\Delta=1$. They differ by the $U(1)_{T'}$ charge, which goes from $-1$ for $\M$ to $1$ for $\a_0$, in steps of $\frac{1}{N-1}$, the $N-1$ operators $\{\M_{SU(N)}\phi^{N-1}\}$ having vanishing $U(1)_{T'}$ charge.

The dressed monopoles of the non-Abelian gauge theory $SU(N)$ with one flavor map to the monopoles of the $U(1)^{N-1}$ Abelian quiver. For instance for $N=4$
\be \label{dressedlinearmap}
\!\!\!\!\!\!\left(\begin{array}{cccc}
 \{\M_{SU(N)}\phi^3\}_1& \{\M_{SU(N)}\phi^4\}_1 & \{\M_{SU(N)}\phi^5\} & \a_0 \\
 \{\M_{SU(N)}\phi^2\}_1 &  \{\M_{SU(N)}\phi^3\}_2 &  \{\M_{SU(N)}\phi^4\}_2 &  \a_1 \\
  \{\M_{SU(N)}\phi\}_1 &   \{\M_{SU(N)}\phi^2\}_2 & \{\M_{SU(N)}\phi^3\}_3 &  \a_2\\
  \M_{SU(N)} & \{\M_{SU(N)}\phi\}_2 & \{\M_{SU(N)}\phi^2\}_3  & 
\end{array}\right)
          \longrightarrow
\left(\begin{array}{cccc}
  \Phi_1 & \M^{1,0,0} & \M^{1,1,0}  & \M^{1,1,1} \\
  \M^{-1,0,0} &  \Phi_2  & \M^{0,1,0}  & \M^{0,1,1}\\
\M^{-1,-1,0} & \M^{-1,0,0} & \Phi_3 & \M^{0,0,1}\\
\M^{-1,-1,-1} & \M^{0,-1,-1}  &  \M^{0,0,-1} &
\end{array}\right)
 \ee
where $\M^{a,b,c}$ are the monopoles of the $U(1)^3$ quiver with topological charges $(a,b,c)$.\footnote{Notice that $\{\M_{SU(N)}\phi^j\}$ are not zero in the chiral ring even if $j\geq N$, we are just using the symbol $\{\M_{SU(N)}\phi^j\}$ to denote the dressed monopole with $j$ factors of $\phi$, of the form \ref{dressedmonform}.} The r.h.s. of eq. (\ref{dressedlinearmap}) are  the Coulomb branch generators of the linear quiver, with scaling dimension $\Delta=1$. The latter operators in turn map to the mesons of the mirror theory, $U(1)$ with $N$ flavors, $\CN=4$.

From eq. (\ref{dressedlinearmap}) we see that the global symmetry $U(1)_{T'}$ of the $SU(N)$ gauge theory descends to the sum of the $N-1$ topological symmetries of the linear quiver. The emergent symmetry which is generated compactifying to $3d$ is enhanced to $SU(N)$. Notice that we are not claiming a precise 1-to-1 map with eq. (\ref{dressedlinearmap}): the global symmetry analysis we made only implies, for instance,  that $\a_0$ is mapped to $\M^{1,1,1}$, and the two dimensional space spanned by $( \{\M_{SU(N)}\phi^5\},\a_1)$ is mapped to the two dimensional space spanned by $( \M^{1,1,0}, \M^{0,1,1})$. It would be interesting to derive the precise mapping of the dressed monopoles to the monopoles of the abelian quiver.\footnote{This mapping allows us to get one more check of the Abelianization duality. We focus on the $SU(2)$ case. Adding to the superpotential term linear in $\a_0$, the meson $tr(\qt q)$ acquires a vev,  breaking the $SU(2)$ gauge symmetry completely. The IR description is a Wess-Zumino model 
\be\label{defa}\W=\b_2 \Tr(\phi^2) =-\frac{\b_2}{2}\det (\phi).\ee 
The Abelianized theory in this case is $U(1)$ with $2$ flavors, using the mapping $\a_0 \leftrightarrow \M_{U(1)}^+$, the linear $\a_0$-deformation corresponds to turning on $\delta\CW=\M_{U(1)}^+$. Taking the mirror dual it becomes an off diagonal mass term 
\be \CW=\Phi(p_1\pt_1 + p_2\pt_2)+p_1\pt_2\ee
We can now integrate out the massive fields, getting a $U(1)$ gauge theory with one flavor and $\CW=-\Phi^2 p_2\pt_1$. Taking the mirror again (using that $U(1)$ with one flavor and $\CW=0$ is dual to the $XYZ$ model), we find a WZ model with $\CW=Z(XY-\Phi^2)$, which is equivalent to (\ref{defa}). (Alternatively, we could have used the monopole duality discussed in the next section). It is interesting that after the linear $\a_0$ deformation, $\b_2$ is not forbidden anymore to acquire a vev.}

On top of these $N^2-1$ generators, that map to the Coulomb branch of the Abelianized $\CN=4$ theory, there are the three operators $\CB,\CM,\tilde{\CB}$ discussed in section \ref{dressed1}.  These three operators in $3d$ have dimension $\Delta[\CB,\tilde{\CB}]=\frac{N}{2}$ and $\Delta[\CM]=1$ and satisfy the same equation $\CB \tilde{\CB}=\CM^N$. They generate the Higgs branch of the Abelianized $\CN=4$ theory, indeed they map to the operators of the $U(1)^{N-1}$ quiver as follows
\bea 
\CB= \varepsilon q (\phi q) \ldots (\phi^{N-1}q) &\longleftrightarrow& \prod_{i=1}^N P_i \\
\label{ccabe}\CM =  \Tr(\qt \phi^{N-1} q) &\longleftrightarrow& P_i\Pt_i \qquad \text{for every}\,\,i \\
\tilde{\CB}= \varepsilon \qt (\qt \phi ) \ldots (\qt \phi^{N-1}) &\longleftrightarrow& \prod_{i=1}^N \Pt_i 
\eea

The $N-1$ $\a_r$'s have vanishing product with $\CB,\CM,\tilde{\CB}$, for the same reasons explained in $4d$. It would be nice to show that also the dressed monopoles have vanishing product with $\CB,\CM,\tilde{\CB}$.  Finding the chiral ring quantum relations satisfied by the dressed monopoles in our $3d$ $\CN=2$ theory is an interesting problem that goes beyond the scope of this paper.

\subsection{$S^3$ partition functions}
\begin{figure}[t]
\begin{center}
\includegraphics[width=7cm]{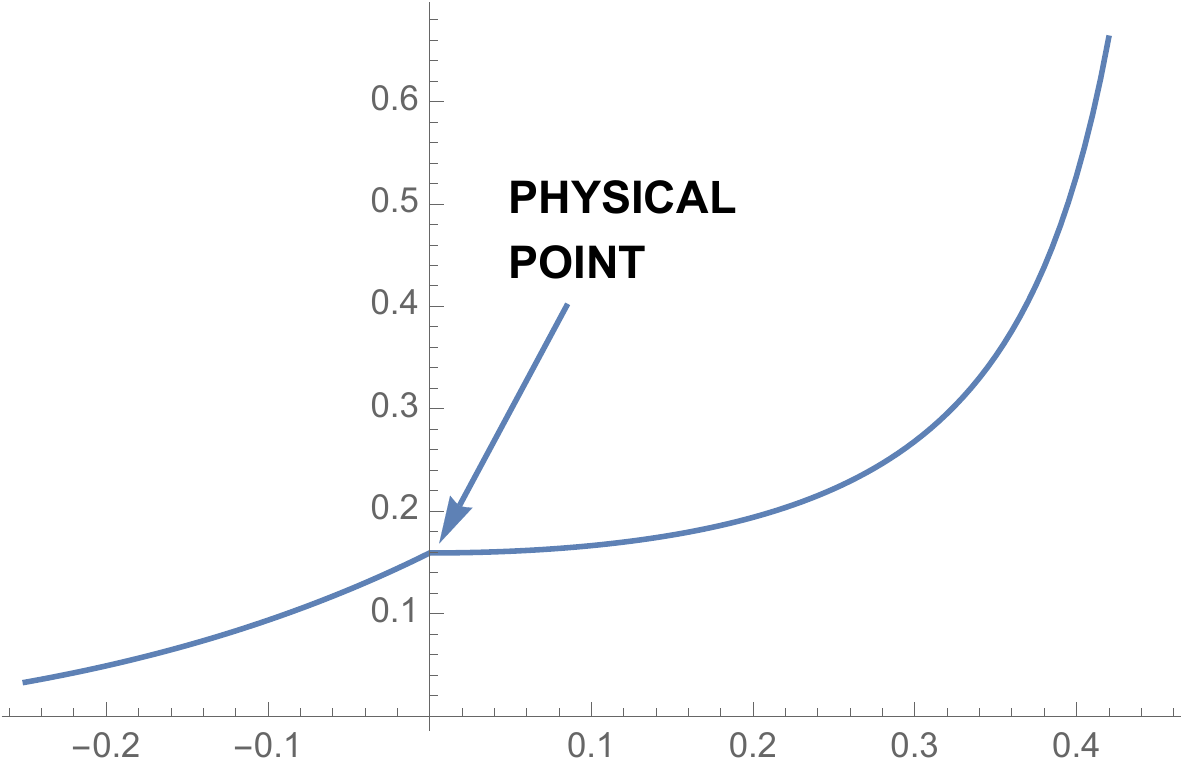} \hspace{1cm} \includegraphics[width=7cm]{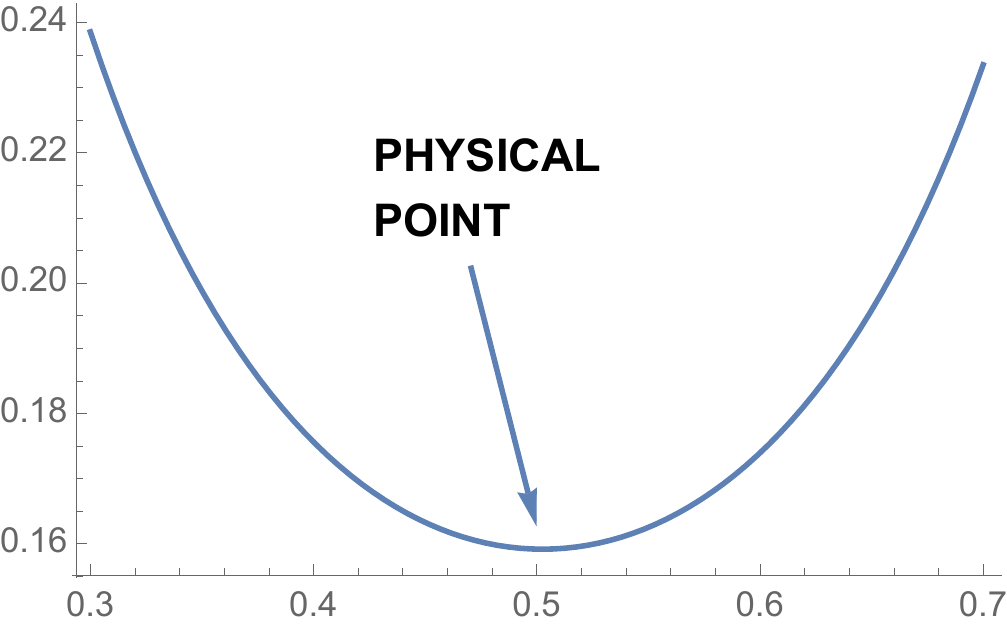}
\caption{On the left $\CZ_{SU(2)}[r_{\phi},\frac{1}{2},0]$. On the right $\CZ_{SU(2)}[0,r_q,0]$. The qualitative behavior for $SU(N)$ is similar.
\label{FIGnumeric}}
\end{center}
\end{figure}

At the level of $S^3$ partition functions, we expect the equality of $\CZ_{SU(N)}[\rf,r_q,b]$ and $\CZ_{U(1)^{N-1}}$ as a function of $3$ variables, for $\rf>0$, which can be checked numerically for small values of $N$. As for the case of $SU(2)$ studied in \cite{Benvenuti:2017lle}, the numerical evaluation of $\CZ_{SU(N)}[\rf,r_q,b]$ present a singularity at $\rf=0$: the first derivative with respect to $\rf$ is discontinuos, as displayed in \ref{FIGnumeric}. We propose that $\CZ_{SU(N)}[\rf,r_q,b]$ should be continued analytically from the region $\rf>0$. 

Using the mapping of the chiral ring generators found in the previous subsection it's possible to find the mapping for the fugacities appearing in $\CZ_{SU(N)}$ and $\CZ_{U(1)^{N-1}}$, which we recall are defined by
\bea \CZ_{SU(N)}[r_{\phi},r_q,b]&=&\nn
 \prod_{r=0}^{N-2}e^{l(1-(2-2r_q-r \cdot \rf))} \prod_{j=2}^Ne^{l(1-(2-j \cdot r_{\phi}))}  \int_{-\infty}^{+\infty}\!\!\frac{\prod_{i>j} (2 sinh(\pi(z_i-z_j)))^2 }{N!} \cdot \\
&& \,\,\cdot \, e^{(N-1) l(1-\rf)} \prod_{i\neq j} e^{l(1-r_{\phi} + i (z_i-z_j))}\prod_i e^{l(1-r_q \pm b - i z_i)} \delta(\sum z_i)dz_i \eea
and
\be \CZ_{U(1)^{N-1}}[r_P,B,\eta_i] = e^{(N-1) l(1-(2-2r_P))} \int_{-\infty}^{+\infty}  \prod_{i=1}^N e^{l(1-r_P \pm B - i (z_i\!-\!z_{i+1}))} e^{2 \pi \sum \eta_j z_j}\prod_{j=1}^{N-1} dz_j \ee

From the mapping of the dressed baryons to the "long mesons" in the quiver 
\be \varepsilon q (\phi q) \ldots (\phi^{N-1}q) \leftrightarrow \prod_{i=1}^N P_i \ee
we can infer where the baryonic and $U(1)_q$ fugacities map:
\be r_P = r_q + \frac{N-1}{2}\rf \ee
and
\be B=b \ee
From the mapping of the dressed monopoles to the monopoles in the quiver (\ref{dressedlinearmap}), we can guess that each Fayet-Iliopoulos fugacity maps to the r-charge of the $SU(N)$ adjoint $\phi$:
\be \eta_i = \rf \ee
Combining these arguments, we arrive at the following equality among the $\CZ_{S^3}$, as functions of three variables:
\be \CZ_{SU(N)}[\rf,r_q,b] = \CZ_{U(1)^{N-1}}[r_P= r_q + \frac{N-1}{2}\rf, B=b , \eta_i=\rf] \ee
We checked this relation numerically for $N=2$ and $N=3$. It only holds if $\rf \geq 0$. $\CZ_{SU(N)}[\rf,r_q,b] $ in the region $\rf<0$ should be analytically continued from the region $\rf>0$. The previous equation provides an analytic continuation in terms of $\CZ_{U(1)^{N-1}}$, which is perfectly regular around its minimum at $r_P=\frac{1}{2},B=\eta_i=0$. In particular the second derivatives, which are the two-point functions of the symmetry currents \cite{Closset:2012vg}, are continuous and positive around the minimum.


\section{Mirror RG flow to $A_{2N-1}$ AD: \emph{Sequential confinement}}\label{SC}

In this section we provide a $3d$ interpretation of the results of \cite{Maruyoshi:2016tqk,Maruyoshi:2016aim} and a further check of the claims of the previous section. The strategy is to reduce the $4d$ RG flow to $3$ dimensions, use $3d$ $\CN=4$ mirror symmetry, and analyze the mirror RG flow applying a duality for $U(N)$ gauge theories with monopole superpotential \cite{Benini:2017dud}.

\subsection{Basic ingredients}\label{basicingr}
Reducing $\CT'_{4d, UV}$ leads to  $3d$ $\CN=4$ theory $SU(N)$ with $2N$ flavors with additional $\CN=2$ superpotential terms and $2(N-1)$ additional chiral $\CN=2$ singlets $\a_r$ and $\b_j$.

In the following we will refer to the mirror of $\CT'_{3d,UV}$ as $\tilde{\CT}_{3d,UV}$. The crucial point is that $\CN=4$ SQCD has a known mirror dual and we can study the RG flow $\tilde{\CT}_{3d,UV} \rightarrow \tilde{\CT}_{3d,IR}$ induced by the additional superpotential terms. In order to proceed we now review the mirror of $U(N)$ with $2N$ flavors, $\CN=4$ susy. We will later adapt those results to the case of $SU(N)$ with $2N$ flavors that we need.

\subsubsection{The mirror of $U(N)$ with $2N$ flavors and the chiral rings map}
The mirror of $U(N)$ with $2N$ flavors is a $U(n_i)$ linear quiver gauge theory \cite{Hanany:1996ie}:
\be \label{UNmirror}
\bpic  \path (-1.5,0) node[circle,draw](x1) {$N$} -- (0,0) node[rectangle,draw](x2) {$2N$};
\draw [-] (x1) to (x2);  \epic \qquad  
   \bpic \draw[<->, thick] (0,0.4) -- (1.6,0.4);  \node[right] at (0.4,1.2){3d}; \node[right] at (0.1,0.7){mirror}; \epic \qquad   
    \bpic  \path (1.5,0) node[circle,draw](x1) {$\,1\,$} -- (2.8,0) node[circle,draw](x2) {$\,2\,$} -- (3.8,0) node(x3) {$\cdots$}-- (5,0) node[circle,draw](x5) {\small$N$} -- (5,1.2) node[rectangle,draw](x6) {$2$}  -- (6,0) node(x8) {$\cdots$} -- (7,0) node[circle,draw](x9) {\small$2$} -- (8,0) node[circle,draw](x10) {\small$1$};
\draw [-] (x1) to (x2);
\draw [-] (x2) to (x3);
\draw [-] (x3) to (x5);
\draw [-] (x5) to (x6);
\draw [-] (x5) to (x8);
\draw [-] (x8) to (x9);
\draw [-] (x9) to (x10);
  \epic 
\ee
The $SU(2N)$ Higgs branch global symmetry on l.h.s. is mapped to the enhanced topological (or Coulomb branch) symmetry $U(1)^{2N-1}\rightarrow SU(2N)_C$ on the r.h.s. The $U(1)$ Coulomb branch symmetry on the l.h.s. is also enhanced to $SU(2)_C$, and is mapped to the $SU(2)$ rotating the $2$ flavors of the central node on r.h.s. A proof of the equality of the refined $\CN=4$ $S^3$ partition functions was given in \cite{Benvenuti:2011ga}. 

The Higgs branch generators of the $U(N)$ with $2N$ flavors $Q_i,\Qt_i$ theory have scaling dimension $\Delta=1$, transform in the adjoint of $SU(2N)$ and map to the Coulomb branch generators of the r.h.s. as follows:
\be \label{HCmap}
\!\!\!\!\!\!\left(\begin{array}{cccc}
  \Tr(Q_1\Qt_1) & \Tr(Q_1\Qt_2)  & \dots & \Tr(Q_1\Qt_{2N}) \\
  \Tr(Q_2\Qt_1) &  \Tr(Q_2\Qt_2) &  \dots &  \Tr(Q_2\Qt_{2N})\\
  \vdots &   \ddots & \ddots &  \vdots\\
 \Tr(Q_{2N}\Qt_1) & \Tr(Q_{2N}\Qt_2) & \ldots  &  \Tr(Q_{2N}\Qt_{2N})
\end{array}\right)
          \leftrightarrow
\left(\begin{array}{ccccc}
  \Tr(\Phi_{U(1)}^L) & \M^{1,0,\ldots,0} & \M^{1,1,\ldots,0} & \dots & \M^{1,1,\ldots,1} \\
  \M^{-1,0,\ldots,0} &  \Tr(\Phi_{U(2)}^L)  & \M^{0,1,0,\ldots,0} & \dots & \M^{0,1,\ldots,1}\\
  \M^{-1,-1,\ldots,0} &  \M^{0,-1,\ldots,0}  & \ddots & \ddots & \vdots\\
  \vdots & \ddots & \ddots & \Tr(\Phi_{U(2)}^R) & \M^{0,\ldots,0,1}\\
\M^{-1,\ldots,-1} & \dots & \ldots &  \M^{0,\ldots,0,-1} &  \Tr(\Phi_{U(1)}^R)
\end{array}\right)
 \ee
 Where $\M^{a_1,a_2,\ldots,a_{2N-1}}$ is the minimal monopole with topological charges $(a_1,a_2,\ldots,a_{2N-1})$ in the r.h.s quiver and the $\Phi$'s are the adjoint of the gauge nodes. See \cite{Benvenuti:2016wet} for discussions and applications of such map.
 
 The Coulomb branch of the $U(N)$ with $2N$ flavors theory is generated by $3N$ operators \cite{Cremonesi:2013lqa}. They transform as $N$ triplets of the global $SU(2)_C$ symmetry, with $\Delta=1,2,\ldots,N$, and map to the Higgs branch of the r.h.s. quiver theory as follows:
\be \label{CHmap} \left(\begin{array}{ccc}
\M^- & \Tr(\phi) & \M^+ \\
 \{\M^-\phi\} & \Tr(\phi^2) & \{\M^+\phi\} \\
     & \ldots  & \\
 \{\M^-\phi^{N-1}\} & \, \Tr(\phi^N) \, & \{\M^+\phi^{N-1}\}
\end{array}\right)
          \longleftrightarrow
\left(\begin{array}{c}
\Tr(q_I\qt^J) \\
 \Tr(q_I p\pt\qt_J) \\
     \ldots \\
    \Tr(q_I p p \ldots \pt \pt \qt_J)   
\end{array}\right)
 \ee
 where $\phi$ is the adjoint in the l.h.s. and $\{M^\pm \phi^j\}$ denote the basic monopole with GNO  charges $(\pm 1,0,0,\ldots,0)$ dressed by $j$ factors of the adjoint field $\phi$. On the r.h.s. $q_I,\qt_J$ denote the $2$ flavors attached to the central $U(N)$ node, $p,\pt$ generically denote the bifundamental fields of the lower row of the quiver.

\subsubsection{Confining $U(N)$ with $N+1$ flavors and $\CW=\M^+$}
The RG flow on the mirror is triggered by linear monopole superpotentials. The analysis is accomplished using a recently found duality (\cite{Benini:2017dud}, section $8$) for  $3d$ $\CN=2$ $U(N_c)$ with $N_f$ flavors (and $N_f$ anti-flavors) and $\CW=\M^+$.  See \cite{Collinucci:2016hpz, Benvenuti:2016wet} for previous examples in the Abelian case and \cite{Amariti:2017gsm} for a brane interpretation. $\M^\pm$ are the basic monopoles of $U(N_c)$ with GNO charges $(\pm 1,0,0,\ldots,0)$. The dual is a Aharony \cite{Aharony:1997gp} magnetic description $U(N_f-N_c-1)$ gauge theory with $N_f$ flavors, $N_f^2+1$ singlets and
\be \CW= \M^- + \M^+ \gamma + \sum \left(M_{N_f}\right)_{ij}\qt_iq_j\ee
The global symmetry on both sides is $U(1)_{\text{topological}} \times SU(N_f)^2 \times U(1)_R$. The duality can be obtained from a real mass deformation of a similar duality for $U(N_c)$ with superpotential $\CW=\M^++\M^-$ \cite{Benini:2017dud}. 

The special case of interest to us is $N_f=N_c+1$, in this case the dual is a Wess-Zumino model with $N_f^2+1$ chiral fields:
\bea U(N_c) , \, N_c+1 \, \text{flavors} \{q_i,\qt_i \} , \,\, \CW=\M^+ \,\,& \longleftrightarrow & \,\, \text{WZ-model} \,\,\, \CW= \gamma_{N_c+1} \, det (M_{N_c+1}) \nn\\
\M^- \,\,\,\,& \longleftrightarrow & \,\, \,\, \gamma_{N_c+1}  \\
\Tr(q_i \qt_j) \,\,\,\,& \longleftrightarrow & \,\, \,\, M_{N_c+1}\nn
\eea
We also displayed the map of the chiral ring generators.

In other words if $N_f=N_c+1$, in the presence of a superpotential $\CW=\M^+$, the gauge theory $U(N_c)$ \textsl{confines}.

We are able to apply the monopole duality since the relevant $U(N_c)$ node loses additional matter like the adjoint field.

\subsection{The general picture of the mirror RG flow}\label{genpic}
Our set of theories of interest can be represented by the following diagram:
\bea \label{DIAGRAM}\nn 
\bpic  \path (-0.8,1.2) node {$\CT'_{3d,UV}$} -- (-1.5,0) node[circle,draw](x1) {$N$} -- (0,0) node[rectangle,draw](x2) {$2N$} -- (-0.8,-1) node {$\CW = \CW_{\CN=4}+\delta \CW_{\CN=2}$}; \node[circle,draw] at (-1.5,0) {$\quad$};
\draw [-] (x1) to (x2);  \epic \qquad  
  & \bpic \draw[<->, thick] (0,0.4) -- (1.6,0.4);  \node[right] at (0.4,1.2){3d}; \node[right] at (0.1,0.7){mirror}; \epic & \qquad   
    \bpic  \path (3,1.2) node {$\tilde{\CT}_{3d,UV}$} (2,0) node[circle,draw](x1) {\small$1$} -- (3,0) node[circle,draw](x2) {\small$2$} -- (4,0) node(x3) {$\cdots$}-- (5,0) node[circle,draw](x5) {\small$\!N\!$} -- (4.5,1) node[circle,draw,red,thick](x6) {\small$1$} -- (5.5,1) node[rectangle,draw](x7) {$1$}  -- (6,0) node(x8) {$\cdots$} -- (7,0) node[circle,draw](x9) {\small$2$} -- (8,0) node[circle,draw](x10) {\small$1$}-- (4.5,-1) node {$\CW = \CW_{\CN=4}+\delta \tilde{\CW}_{\CN=2}$};
\draw [-] (x1) to (x2);
\draw [-] (x2) to (x3);
\draw [-] (x3) to (x5);
\draw [-] (x5) to (x6);
\draw [-] (x5) to (x7);
\draw [-] (x5) to (x8);
\draw [-] (x8) to (x9);
\draw [-] (x9) to (x10);
  \epic \nn \\
\qquad\qquad \bpic \draw[->,thick] (0,1) -- (0,-2); \node[right] at (0.3,0) {RG flow:}; \node[right] at (0.3,-0.7) {matter fields}; \node[right] at (0.3,-1.4) {integrated out};\epic &&\qquad \quad\qquad\qquad \bpic \draw[->,thick] (0,1) -- (0,-2); \node[right] at (0.3,0) {RG flow:}; \node[right] at (0.3,-1) {gauge nodes confine};\epic\\
\CT'_{3d,IR} \qquad  \qquad \qquad
  & \bpic \draw[<->, thick] (0,0.4) -- (1.6,0.4);  \node[right] at (0.4,1.2){3d}; \node[right] at (0.1,0.7){mirror}; \epic &\qquad   \qquad
\bpic  
\node[right] at (1,1){$\tilde{\CT}_{3d,IR}$:};
\node[circle,draw,red,thick](x1) at (0,0){$1$};
\node[rectangle,draw](x2) at (1.5,0){$N$};    
\node[right] at (3,0){$\CW = \CW_{\CN=4}$};
\draw[-] (x1) -- (x2);
  \epic \nn
\eea
where we are already anticipating the result: on the r.h.s. in the IR the manifestly $\CN=4$ theory $U(1)$ with $N$ flavors appears, as expected.

The upper part of \ref{DIAGRAM} is obtained from \ref{UNmirror}, where on the l.h.s. we gauged the $U(1) \subset SU(2)$ topological symmetry, so the gauge group, from $U(N)$, becomes $SU(N)$ and the $SU(2)$ topological symmetry is replaced by a $U(1)_{baryonic}$ symmetry. On the r.h.s. this maps to gauging one of the two flavors (red node), breaking the global $SU(2)$ symmetry and gaining an additional $U(1)$ topological symmetry.

$\delta \tilde{\CW}_{\CN=2}$ is given by the mirror of 
\be \label{deltaW} \delta \CW_{\CN=2} =  \sum_{i=1}^{2N-1} \Tr(\qt_i q^{i+1}) + \sum_{r=0}^{N-2} \a_r \sum_{i=0}^r \Tr(\qt_{2N-r+i} q^{i+1}) + \sum_{j=2}^{N} \b_j\, \Tr(\phi^j) \ee
The mirror of $\delta \CW_{\CN=2}$ can be worked out adapting the maps (\ref{HCmap}) and (\ref{CHmap}) from $U(N)$ to $SU(N)$. 

According to (\ref{HCmap}), the first sum in (\ref{deltaW})  is mapped to a term linear in the $2N-1$ monopoles with precisely one positive topological charge:
\be \M^{1,0,\ldots,0} + \M^{0,1,0\ldots,0} + \ldots + \M^{0,\ldots,0,1} \ee
There are linear monopole superpotential only for the nodes in the lower row of the quiver $\tilde{\CT}_{3d,UV}$, the upper $U(1)$ gauge node attached to the central $U(N)$ node will never have monopole potentials.

Using (\ref{HCmap}) again, the second sum in (\ref{deltaW}) is mapped to flipping terms for monopoles with negative topological charges
\be \a_0\M^{-1,\ldots,-1} + \a_1 (\M^{-1,\ldots,-1,0} +\M^{0,-1,\ldots,-1} ) +\ldots + \a_{N-2} (\M^{-1,\ldots,-1,0,\ldots,0} + \ldots) \ee

Finally, the third sum in (\ref{deltaW}), $\sum_{j=2}^{N} \b_j \Tr(\phi^j)$, is mapped to flipping terms for mesonic operators appearing in the r.h.s. of the map (\ref{CHmap}), adapted from $U(N)$ to $SU(N)$ gauge symmetry.

\subsubsection*{Sequential confinement}
We start applying the monopole duality to the leftmost $U(1)$ node in the upper-right quiver in \ref{DIAGRAM}. The $U(1)$ node confines and the Seiberg dual mesons give mass to the adjoint of the close-by $U(2)$ node. At this point the $U(2)$ node has no adjoint, $3$ flavors and a monopole superpotential $\M^+$, so we apply the monopole duality to the $U(2)$ node. 

This pattern goes on until the left tail has disappeared and we reach the central node $U(N)$. When the central node confines, some of the Seiberg dual mesons give mass to the adjoint of the $U(N-1)$ node, some become bifundamental fields for the $3$ groups $U(1) \times U(1)_F \times U(N-1)$. Going down along the right tail, at each dualization step we generate one more bifundamental flavor between the two upper nodes. At the end we are left with just $U(1) \times U(1)_F$ with $N$ bifundamental hypers, that is $U(1)$ gauge theory with $N$ hypermultiplet flavors.

This is the qualitative story, in the following we analyze in detail the process of \emph{sequential confinement} including the superpotential, and confirm that the RG flow lands on $U(1)$ with $N$ flavors with $\CN=4$ supersymmetry. The only gauge-singlet, among the $N\!-\!1$ $\a_r$'s, the $N\!-\!1$ $\b_j$'s and the $2N\!-\!1$ $\gamma_n$'s, that is massless in the IR is the $\gamma_{N+1}$ singlet, generated when dualizing the central $U(N)$ node into a Wess-Zumino $\CW=\gamma_{N+1} det(M_{N+1})$. $\gamma_{N+1}$ sits in the $\CN=4$ vector multiplet of the $U(1)$ gauge theory. The IR superpotential, modulo a sign, is
\be \CW_{\tilde{\CT}_{IR}} = \gamma_{N+1} \sum_{i=1}^N \Qt_i Q_{N-i+1} \ee
where $Q_i,\Qt_i$ is the fundamental hypermultiplet generated at the $i^{th}$-step, dualizing down the second tail.

Let us make a final comment: if we had considered a non maximal Jordan block (and also non next-to-maximal), the sequential confinement would have stopped before, and the IR mirror theory would contain a non-Abelian node without the adjoint, so the mirror would clearly be only $\CN=2$ supersymmetric.


\subsection{Mirror RG flow to $A_{3}$ AD: the superpotential}\label{su2SC}
Since including the analysis of the superpotential leads to complicated expressions, we focus first on the cases $N=2,3$. We will later comment about the generalization to $N>3$.

We start from $\CN=4$ $SU(2)$ SQCD with four flavors, whose mirror is the $\CN=4$ quiver \cite{Hanany:1996ie, Benini:2010uu,Benvenuti:2011ga}
\be\label{sqcd2}
\begin{tikzpicture}[->, thick]
\node[shape=circle, draw] (1) at (0,0) {$1$};
\node[shape=circle, draw] (2) at (1,1) {$2$};
\node[shape=circle, draw] (3) at (2,0) {$1$};
\node[shape=circle, draw, red] (4) at (0,2) {$1$};  
\node[shape=rectangle, draw] (5) at (2,2) {$\,1\,$};
\node[] (6) [left= .1cm of 1] {$1$};
\node[] (7) [left= .1cm of 4] {$2$};
\node[] (8) [right= .1cm of 3] {$3$};
\node[] (9) [right= .1cm of 5] {$4$};
\draw[-] (1) -- (2);\draw[-] (2) -- (3);\draw[-] (2) -- (4);\draw[-] (5) -- (2);
\end{tikzpicture}
\ee
We numbered the abelian groups in the quiver and we call $p_i,\tilde{p}_i$ the $U(2)\times U(1)_i$ bifundamentals. The cartan subgroup of the $SO(8)$ global symmetry of the theory is identified with the topological symmetries of the four gauge nodes. We are only interested in the $SU(4)$ symmetry associated with the nodes $U(1)_1$, $U(1)_3$ and $U(2)$. The singlets in the abelian vector multiplets will be denoted $\varphi_i$ ($i=1,2,3$) whereas the trace and traceless parts of the $U(2)$ adjoint are $\hat{\phi}_2$ and $\phi_2$ respectively. The operators $\Tr\phi^2$ and the monopole of $SU(2)$ SQCD are mapped on the mirror side to $\tilde{p}_4p_3\tilde{p}_3p_4$ and $\tilde{p}_4p_3\tilde{p}_3p_4+\tilde{p}_2p_3\tilde{p}_3p_2$ respectively.

The $SO(8)$ global symmetry of SQCD arises quantum mechanically in the mirror theory, due to the presence of monopole operators of scaling dimension $1$, whose multiplets contain conserved currents \cite{Gaiotto:2008ak}. We recall that the map between off-diagonal components of the $SU(4)$ meson and monopoles is as follows: 
\be\label{meson}\left(\!\!\begin{array}{cccc}
 & \tilde{q}_1q^2 & \tilde{q}_1q^3 & \tilde{q}_1q^4\\
 \tilde{q}_2q^1 & & \tilde{q}_2q^3 & \tilde{q}_2q^4\\
 \tilde{q}_3q^1 & \tilde{q}_3q^2 & & \tilde{q}_3q^4\\ 
 \tilde{q}_4q^1 & \tilde{q}_4q^2 & \tilde{q}_4q^3 & \\
\end{array}\!\!\right)
\!\!\!\leftrightarrow\!\!\! \left(\!\!\begin{array}{cccc}
 & \M^{+00} & \M^{++0} & \M^{+++} \\
 \M^{-00} & & \M^{0+0} & \M^{0++} \\
 \M^{--0} & \M^{0-0} & & \M^{00+}\\
 \M^{---} & \M^{0--} & \M^{00-} & \\
\end{array}\!\!\right)\ee
The Cartan components of the meson matrix are mapped to $\varphi_1$, $\varphi_3$ and $\hat{\phi}_2$.
In (\ref{meson}) we have included only the charges under the topological symmetries related to $U(1)_1$, $U(1)_3$ and $U(2)$, the others being trivial.

Mapping the deformations of $\CN=4$ $SU(2)$ with $4$ flavors to the mirror theory $\tilde{\CT}'_{UV}$, we find that the mirror RG flow starts from
\bea\label{mirror2}\W_{\tilde{\CT}'_{UV}}=\sum_i\varphi_i\tilde{p}_ip^i-\hat{\phi}_2(\sum_i\tilde{p}_ip^i)-\Tr(\phi_2(\sum_ip^i\tilde{p}_i))+\nn\\ +\M^{+00}+\M^{0+0}+\M^{00+}+\alpha_0\M^{---}+\beta_2\tilde{p}_4p_3\tilde{p}_3p_4.\qquad\eea
According to the monopole duality, the gauge group $U(1)_1$ confines 
\be\label{sqcd2p}
\begin{tikzpicture}[->, thick]
\node[shape=circle, draw] (2) at (1,1) {$2$};
\node[shape=circle, draw] (3) at (2,0) {$1$};
\node[shape=circle, draw, red] (4) at (0,2) {$1$};  
\node[shape=rectangle, draw] (5) at (2,2) {$\,1\,$};
\node[] (7) [left= .1cm of 4] {$2$};
\node[] (8) [right= .1cm of 3] {$3$};
\node[] (9) [right= .1cm of 5] {$4$};
\draw[-] (2) -- (3);\draw[-] (2) -- (4);\draw[-] (5) -- (2);
\end{tikzpicture}
\ee
leaving behind the $U(2)$ adjoint chiral $M_2$, which enters in the superpotential with terms 
$$\gamma_2\det M_2+\varphi_1\Tr M_2-\hat{\phi}_2\left(\Tr M_2+\sum_{i>1}\tilde{p}_ip^i\right)- \nn \\ \Tr\left[\phi_2\!\left(M_2 +\sum_{i>1}p^i\tilde{p}_i\right)\!\right]$$
 $M_2$ and $\phi_2$ become massive and can be integrated out, the equations of motion impose the constraint $M_2=-\sum_{i>1}p^i\tilde{p}_i$.

At this stage the $U(2)$ gauge group has three flavors and no adjoint matter, so according to the monopole duality it confines and is traded for a $3\times3$ chiral multiplet $M_3$, which is nothing but the dual of $\tilde{p}_ip_j$ ($i,j=2,3,4$). This also generates the superpotential term $\gamma_3\det M_3$.  The constraint $M_2=-\sum_{i>1}p^i\tilde{p}_i$ allows to express $\det M_2$ in terms of traces of $M_3$: 
\be\label{id2}\!\!\!\det\! M_2\!=\!\frac{(\Tr M_2)^2-\Tr M_2^2}{2}\!=\!\frac{(\tilde{p}_ip^i)^2-\Tr((\tilde{p}_ip_j)^2)}{2}\!=\!\frac{(\Tr M_3)^2-\Tr M_3^2}{2}.\ee

In theory (\ref{sqcd2}) the cartan subgroup of the $U(3)$ symmetry under which $\tilde{p}_ip_j$ ($i,j=2,3,4$) transforms in the adjoint representation is gauged: the $U(1)_{2,3,4}$ symmetries are generated respectively by the $3\times3$ matrices $\text{diag}(1,0,0)$, $\text{diag}(0,0,1)$ and $\text{diag}(0,1,0)$. Our convention is that these groups act in the same way on the matrix $M_3$ after confinement of the $U(2)$ gauge group. As a result, the off-diagonal components of $M_3$ become bifundamental hypermultiplets charged under the leftover $U(1)_i$ symmetries and we relabel the fields as follows:
$$(M_3)_1^2,(M_3)_2^1\!\leftrightarrow\!Q_1,\tilde{Q}_1;\,\, (M_3)_1^3,(M_3)_3^1\!\leftrightarrow\!v,\tilde{v};\,\, (M_3)_2^3,(M_3)_3^2\!\leftrightarrow\!w,\tilde{w}.$$
After confinement of the $U(2)$ gauge group the theory (\ref{sqcd2}) becomes:
\be \label{step1} \bpic [->, thick] 
\node[shape=circle, draw, minimum height=.9cm] (1) at (0,-1) {$1$};
\node[shape=rectangle, draw, minimum height=.75cm] (2) at (1.5,1) {$\;1\;$};
\node[shape=circle, draw, red, minimum height=.9cm] (3) at (-1.5,1) {$1$};
\node[] (4) [left= .2cm of 3] {$2$};
\node[] (5) [right= .2cm of 1] {$3$};
\node[] (6) [right= .2cm of 2] {$4$};
\node[] (7) at (0,1.3) {$Q_1,\tilde{Q}_1$};
\node[] (8) at (-1.5,0) {$v,\tilde{v}$};
\node[] (9) at (1.5,0) {$w,\tilde{w}$};
\draw[-] (1) -- (2);\draw[-] (1) -- (3);\draw[-] (2) -- (3); \epic \ee

The fields $\varphi_i$ now appear only in the superpotential terms 
\be\W=(\varphi_2-\hat{\phi}_2)(M_3)_1^1+\hat{\phi}_2(M_3)_2^2+(\varphi_3-\hat{\phi}_2)(M_3)_3^3\dots\ee 
As a consequence they become massive and their F-terms set to zero the diagonal components of $M_3$. The remaining fields are $\alpha_0$, $\beta_2$, $\gamma_{2,3}$ and the three bifundamental hypermultiplets with superpotential 
\be\label{mirror23}\W=-\frac{\gamma_2}{2}(\tilde{Q}_1Q_1+\tilde{v}v+\tilde{w}w)+\beta_2(\tilde{w}w)+\gamma_3(\tilde{Q}_1\tilde{v}w+\tilde{w}vQ_1)+\M^++\alpha_0\M^-,\ee 
where the monopoles are charged under the topological symmetry of $U(1)_3$. 

Finally, the gauge group $U(1)_3$ confines and its meson components $\tilde{v}v$, $\tilde{w}w$ and $\tilde{v}w$, $\tilde{w}v$ become elementary fields of the theory.  The first two are singlets, which we call $x$ and $y$, whereas the other two are charged under the surviving gauge group $U(1)_2$ and we call them $Q_2$, $\tilde{Q}_2$.

\be \bpic [->, thick]
\node[shape=circle, draw, red, minimum height=.9cm] (19) at (4.5,0) {$1$};
\node[shape=rectangle, draw, minimum height=.75cm] (20) at (6.5,0) {$\;1\;$};
\node[] (22) [left= .2cm of 19] {$2$};
\node[] (23) [right= .2cm of 20] {$4$};

\draw[-] (6.12,0.1) -- (4.96,0.1);
\draw[-] (6.12,-0.1) -- (4.96,-0.1);

\node[right] at (4.8,0.4) {$Q_1,\!\tilde{Q}_1$};
\node[right] at (4.8,-0.4) {$Q_2,\!\tilde{Q}_2$};
\epic \ee
 After confinement of $U(1)_3$ (\ref{mirror23}) becomes
\be\label{mirror24}\W=-\frac{\gamma_2}{2}(\tilde{Q}_1Q_1+x+y)+\beta_2y+\gamma_3(\tilde{Q}_1Q_2+\tilde{Q}_2Q_1) +\gamma_2'(xy-\tilde{Q}_2Q_2)+\alpha_0\gamma_2',\ee 
and all the fields except $\gamma_3$ and the two $U(1)_2$ flavors become massive. Integrating them out we are left with 
\be\W_{\tilde{\CT}'_{IR}}=\gamma_3(\tilde{Q}_1Q_2+\tilde{Q}_2Q_1),\ee
which is equivalent to the standard superpotential of $\CN=4$ SQED with two flavors after a change of variable and this is precisely the mirror of $A_3$ Argyres-Douglas theory proposed in \cite{Nanopoulos:2010bv}.

\subsection{Mirror RG flow to $A_{5}$ AD: the superpotential}\label{su3mm}

We now focus on the case $N=3$. The prescription to obtain the $A_5$ AD theory is to start from $SU(3)$ SQCD with six flavors, turn on five off-diagonal mass terms and flip the operators $\Tr\Phi^2$ and $\Tr\Phi^3$. We also introduce the two flipping fields ($\a_0$ and $\a_1$) which do not decouple in the IR. The superpotential is 
\be\label{uva5}\W_{\CT'_{3d, UV}}=\sum_{i=1}^6\Tr(\tilde{q}_i\phi q^i)+\beta_2\Tr(\phi^2)+\beta_3\Tr(\phi^3)+\sum_{i=1}^5\Tr(\tilde{q}_iq^{i+1}) +\alpha_{0}\Tr(\tilde{q}_6q^1)+\alpha_1\Tr(\tilde{q}_5q^1+\tilde{q}_6q^2).\ee 
We refer to this model as the $\CT'_{3d, UV}$ theory. 
Its mirror is the quiver \cite{Gaiotto:2008ak,Benini:2010uu}
\be \label{sqcd3} \bpic[->, thick]
  \path (2,1.4) node {$\tilde{\CT}_{3d,UV}$} (1,0) node[circle,draw](x1) {$1$} -- (3,0) node[circle,draw](x2) {$2$} -- (5,0) node[circle,draw](x5) {$3$} -- (4,1.5) node[circle,draw,red,thick](x6) {$1$} -- (6,1.5) node[rectangle,draw,thick](x7) {$1$}  --  (7,0) node[circle,draw](x9) {$2$} -- (9,0) node[circle,draw](x10) {$1$}; 
\draw [-] (x1) to (x2);
\draw [-] (x2) to (x5);
\draw [-] (x5) to (x6);
\draw [-] (x5) to (x7);
\draw [-] (x5) to (x9);
\draw [-] (x9) to (x10);
\node at (3.8,0.9) {$q_1,\qt_1$};
\node at (6.2,0.9) {$q_2,\qt_2$};
\node at (6,-0.3) {$p_3,\pt_3$};
\node at (8,-0.3) {$p_4,\pt_4$};
\node at (4,-0.3) {$p_2,\pt_2$};
\node at (2,-0.3) {$p_1,\pt_1$};
  \epic
\ee 


Every unitary gauge group gives rise to a topological $U(1)$ symmetry and the $U(1)^5$ global symmetry arising from the nodes in the lower row enhances to $SU(6)$. 

We denote the gauged $U(1)$ in the upper-row (depicted in red) $U(1)_{\text{red}}$, and the flavor $U(1)$ as $U(1)_F$. These two nodes will survive in the IR. We denote the bifundamental matter fields in the quiver as explained in (\ref{sqcd3})\footnote{We slightly change notation with respect to the $SU(2)$ case since the four tails are not on equal footing for $N>2$.}. Since the theory is $\CN=4$, every vector multiplet includes a chiral multiplet transforming in the adjoint representation. We denote the trace part of the adjoint chirals in the two tails (from left to right) as $\varphi_i$ (i=1,...,5), the singlet of the $U(1)_{\text{red}}$ as $\varphi_6$ and the traceless part for the non abelian nodes as $\phi_{2,L}$, $\phi_{2,R}$ and $\phi_3$. 

In order to study the mirror RG flow we need to map in the mirror theory (\ref{sqcd3}) all the superpotential terms appearing in (\ref{uva5}).  Adapting the mapping (\ref{CHmap}) from $U(N)$ to $SU(N)$, we claim that the Casimirs $\Tr(\phi^2)$ and $\Tr(\phi^3)$ of the UV $\CN=4$ SU(3) SQCD are mapped in the mirror (\ref{sqcd3}) to
\be\label{mircb}\Tr (\phi^2)\leftrightarrow q_2p_3\tilde{p}_3\tilde{q}_2;\quad \Tr (\phi^3)\leftrightarrow q_2p_3p_4\tilde{p}_4\tilde{p}_3\tilde{q}_2.\ee
Using (\ref{mircb}) and the observations of the section \ref{genpic} the complete UV mirror superpotential reads
\be\label{mirror3}\begin{array}{ll}\W_{\tilde{\CT}_{3d,UV}}=&\W_{\CN=4}+\sum_{i=1}^5 \M_i^+ + \alpha_1(\M^{----0}+\M^{0----}) + \alpha_0\M^{-----} + \\ 
&\beta_2q_2p_3\tilde{p}_3\tilde{q}_2+\beta_3q_2p_3p_4\tilde{p}_4\tilde{p}_3\tilde{q}_2.\end{array}\ee 
where $\M^\pm_i$ are the $5$ monopoles with just one lower-row topological charge turned on.

The rest of this section is devoted to the study of the RG flow. As in section \ref{su2SC}, our basic tool is the monopole duality for $\CN=2$ $U(N)$ SQCD with $N+1$ flavors (and no adjoint matter) reviewed in section \ref{basicingr}. Using this duality, the final result will be that all the gauge nodes at which we have turned on the monopole superpotential term $\M^+$ confine and our strategy is to follow the evolution of theory (\ref{mirror3}) step-by-step, sequentially dualizing one node at each step. This is essentially the mirror counterpart of integrating out massive flavors one by one. At the end of this process, once we have dualized all nodes with the monopole term, the two monopoles multiplying $\alpha_1$ in (\ref{mirror3}) become the same operator, in analogy with the mirror theory (\ref{uva5}), where both $\Tr(\tilde{q}_5q^1)$ and $\Tr(\tilde{q}_6q^2)$ become $\Tr(\qt \phi q)$ in the IR.

The first step is to apply the monopole duality to the abelian node on the left, the relevant superpotential terms are 
\be\W= \M_1^+ + \varphi_1\tilde{p}_1p_1 +  \varphi_2(\Tr\tilde{p}_2p_2-\tilde{p}_1p_1)+\Tr\phi_{2,L}(\tilde{p}_2p_2-p_1\tilde{p}_1)+\M_2^+ +\dots\ee 
The theory becomes
\be \bpic[->, thick]
  \path  (3,0) node[circle,draw](x2) {$2$} -- (5,0) node[circle,draw](x5) {$3$} -- (4,1.5) node[circle,draw,red,thick](x6) {$1$} -- (6,1.5) node[rectangle,draw,thick](x7) {$1$}  --  (7,0) node[circle,draw](x9) {$2$} -- (9,0) node[circle,draw](x10) {$1$}; 
\draw [-] (x2) to (x5);
\draw [-] (x5) to (x6);
\draw [-] (x5) to (x7);
\draw [-] (x5) to (x9);
\draw [-] (x9) to (x10);
\node at (3.8,0.9) {$q_1,\qt_1$};
\node at (6.2,0.9) {$q_2,\qt_2$};
\node at (6,-0.3) {$p_3,\pt_3$};
\node at (8,-0.3) {$p_4,\pt_4$};
\node at (4,-0.3) {$p_2,\pt_2$};
  \epic
\ee 
The $2\times2$ chiral $M_2$ appears, and the above superpotential terms become: 
\be\label{dual1}\W=\gamma_2\det M_2 + \varphi_1\Tr M_2 + \varphi_2(\Tr\tilde{p}_2p_2-\Tr M_2)+\Tr\phi_{2,L}(\tilde{p}_2p_2-M_2)+\M_2^+ +\dots\ee 
The fields $M_2$, $\varphi_2$ and $\phi_{2,L}$ are now massive and can be integrated out. From the above formula one can easily see that the equations of motion identify $M_2$ with the $2 \times 2$ matrix $p_2 \pt_2$.

At this stage the neighbouring $U(2)$ node has three flavors and no adjoint multiplets ($\phi_{2,L}$ has become massive), so the left $U(2)$ node confines and gets replaced by a $3 \times 3$ chiral multiplet $M_3$ and singlet $\gamma_3$: 
\be \label{firststep}\bpic[->, thick]
  \path  (5,0) node[circle,draw](x5) {$3$} -- (4,1.5) node[circle,draw,red,thick](x6) {$1$} -- (6,1.5) node[rectangle,draw,thick](x7) {$1$}  --  (7,0) node[circle,draw](x9) {$2$} -- (9,0) node[circle,draw](x10) {$1$}; 
\draw [-] (x5) to (x6);
\draw [-] (x5) to (x7);
\draw [-] (x5) to (x9);
\draw [-] (x9) to (x10);
\node at (3.8,0.9) {$q_1,\qt_1$};
\node at (6.2,0.9) {$q_2,\qt_2$};
\node at (6,-0.3) {$p_3,\pt_3$};
\node at (8,-0.3) {$p_4,\pt_4$};  \epic \ee 
 Integrating out all massive fields we get 
\be\begin{array}{ll}\W=&\gamma_2 \det (p_2 \pt_2) +\gamma_3\det M_3 +\varphi_2(\Tr p_2\tilde{p}_2)+\varphi_3(\sum_{i=1,2} \Tr\tilde{q}_iq^i+\Tr\tilde{p}_3p_3-\Tr p_2\tilde{p}_2)+ \label{W321}\\ 
& \phi_3 \Tr(\sum_{i=1,2} \tilde{q}_iq^i+\tilde{p}_3p_3-p_2\tilde{p}_2)+  \dots\end{array}\ee 

As before, the multiplets $M_3$, $\varphi_3$ and $\phi_3$ become massive and can be integrated out, implying that the $U(3)$ gauge group now has four flavors and no adjoint matter.

Let us now pause to explain how to treat the determinants which arise dynamically at each dualization step, like $\gamma_2 \det (p_2 \pt_2) +\gamma_3\det M_3$ in (\ref{W321}). We need to rewrite these determinants in terms of the fields that survive the various dualization steps as in the $SU(2)$ case discussed before.

Considering for instance the term $\gamma_2\det M_2$ generated at the first step, as already explained F-terms identify the multiplet $M_2$ with $\tilde{p}_2p_2$ and then when the $U(2)_L$ node confines $p_2\tilde{p}_2$ is identified with $M_3$. So we need to rewrite $\det M_2$ in terms of the surviving field $M_3$. This is accomplished by first rewriting the determinants in terms of traces \footnote{We will use the following identities for $k\times k$ matrices $M_k$: 
\be\label{id2x2}\det M_2=\frac{(\Tr M_2)^2}{2}-\frac{\Tr M_2^2}{2},\ee
\be\label{id3x3}\det M_3=\frac{\Tr M_3^3}{3}+\frac{(\Tr M_3)^3}{6}-\frac{\Tr M_3}{2}\Tr M_3^2,\ee
\be\label{id4x4}\det M_4=-\frac{\Tr M_4^4}{4}+\frac{(\Tr M_4)^4}{24}+\frac{(\Tr M_4^2)^2}{8}+\frac{\Tr M_4}{3}\Tr M_4^3-\frac{(\Tr M_4)^2}{4}\Tr M_4^2.\ee 
These are special cases of the formula 
\be\label{iddet}\det M_k=\sum_{n_1,\dots,n_k}\prod_{l=1}^k\frac{(-1)^{n_l+1}}{l^{n_l}n_l!}(\Tr M_k^l)^{n_l},\ee 
where the sum is taken over the set of all integers $n_l\geq0$ satisfying the relation $\sum_{l=1}^k ln_l=k.$

These identities will be used to handle the superpotential terms which arise dynamically. 

We will also need the following simple observation: in the mirror quiver there are bifundamental hypermultiplets $b_i^j,\tilde{b}_i^j$ charged under $U(k)\times U(k+1)$. The operator $M_k=\tilde{b}_i^kb_k^j$ transforms in the adjoint of $U(k)$ whereas $M_{k+1}=b_i^k\tilde{b}_k^j$ transforms in the adjoint of $U(k+1)$ and the following identity holds \be\label{bifid}\Tr M_k^n=\Tr M_{k+1}^n\quad\forall n.\ee
}. Using (\ref{id2x2}) and 
(\ref{bifid}) the relation is 
$$\det M_2=\frac{1}{2}((\Tr\tilde{p}_2p_2)^2-\Tr(\tilde{p}_2p_2)^2)=\frac{1}{2}((\Tr M_3)^2-\Tr M_3^2).$$ 
More in general, if we had considered the mirror of $SU(N)$ SQCD with 2N flavors, by turning on monopole superpotential terms at all the nodes along a tail the various nodes confine and at the k-th step the $U(k)$ gauge group disappears and is replaced by a $(k+1)\times(k+1)$ chiral multiplet $M_{k+1}$. The superpotential term $\gamma_2\det M_2$ generated at the first step can be rewritten as 
\be \frac{\gamma_2}{2}((\Tr M_{k+1})^2-\Tr M_{k+1}^2).\ee
A similar observation applies to the terms generated at the subsequent dualization steps, using (\ref{id3x3}), (\ref{id4x4}) and generalizations thereof. In this way it is possible to keep track of all the terms generated along the process of sequential confinement and write all the superpotential terms as functions of the surviving fundamental fields of the theory.  

Going back to the analysis of our RG flow, we can now dualize the $U(3)$ node of (\ref{firststep}), generating the superpotential term $\gamma_4\det M_4$. Now the operators $\tilde{q}_1q_1$ and $\tilde{q}_2q_2$ become diagonal elements of $M_4$, which are elementary fields of the theory. The chirals $\varphi_3$ and $\varphi_6$ become massive and their F-terms set to zero the two diagonal elements of $M_4$ they couple to. Another important fact is that, since a $U(2)\times U(1)$ subgroup of $SU(4)$ is gauged in the quiver, the massless components of $M_4$ decompose as a $U(2)$ adjoint (which we call $\Psi$), a $U(2)\times U(1)_{\text{red}}$ bifundamental, a $U(2)\times U(1)_F$ bifundamental (we denote them as $v,\tilde{v}$ and $w,\tilde{w}$ respectively) and a  $U(1)_{\text{red}}\times U(1)_F$ bifundamental which we call $Q_1,\Qt_1$:
\be\label{expr4} M_4=\left(\begin{array}{ccc}
0 & Q_1 & \tilde{v} \\
\Qt_1 & 0 & \tilde{w} \\
v & w & \Psi
        \end{array}\right)
\ee 

 All in all, we get the theory 
\be \label{step3}
 \begin{tikzpicture}[->, thick]
\node[shape=circle, draw, minimum height=.8cm] (1) at (4,0) {$1$};
\node[shape=circle, draw, minimum height=.8cm] (2) at (2,0) {$2$};
\node[shape=circle, draw, minimum height=.8cm,red,thick] (3) at (0.9,1.5) {$1$};
\node[shape=rectangle, draw, minimum height=.8cm,thick] (4) at (3.1,1.5) {$\,\,1\,\,$};
\node[] (5) at (3,-0.3) {$p_4,\tilde{p}_4$};
\node[] (6) at (0.9,0.7) {$v,\tilde{v}$}; 
\node[] (7) at (3.1,0.7) {$w,\tilde{w}$}; 
\node[] (8) at (2,1.8) {$Q_1\!,\!\Qt_1$};
\draw[-] (1) -- (2);
\draw[-] (2) -- (3);
\draw[-] (2) -- (4);
\draw[-] (4) -- (3);
\end{tikzpicture}
\ee

\noindent with the following, complete, superpotential 
\be\label{a5potential}\begin{array}{ll}\W=&\gamma_2(..)+\gamma_3(..)+\gamma_4\det M_4+\varphi_4(\Tr\Psi-\tilde{p}_4p_4)+\varphi_5\tilde{p}_4p_4 + \Tr\phi_{2,R}(p_4\tilde{p}_4-\Psi)+\\ & \M_4^+ + \M_5^+ + \alpha_1(\M_4^- + M') + \alpha_0\M^{--} + \beta_2\tilde{w}w + \beta_3\Tr(wp_4\tilde{p}_4\tilde{w}),\end{array}\ee 
where $\M^{\pm}_{4,5}$ are the monopole operators charged under one of the topological symmetries of the $U(2)$ and $U(1)$ nodes of the lower row in (\ref{step3}). $\M^{--}$ is the monopole with charge -1 under both topological symmetries and $M'$ is the operator to which $\M^{0----}$ (appearing in (\ref{mirror3})) is mapped under these dualities. We will discuss it in more in detail later. 

$\Psi$, $\varphi_4$ and $\phi_{2,R}$ become massive, leaving
\be\label{expr4prime} M_4=\left(\begin{array}{ccc}
0 & Q_1 & \tilde{v} \\
\Qt_1 & 0 & \tilde{w} \\
v & w & p_4\tilde{p}_4
        \end{array}\right)
\ee 
and
\be\label{a5potentialprime}\begin{array}{ll}\W=&\gamma_2(..)+\gamma_3(..)+\gamma_4\det M_4+\varphi_5\tilde{p}_4p_4 + \M_4^+ + \M_5^+ +\\ &+ \alpha_0\M^{--}  +  \alpha_1(\M_4^- + M') + \beta_2\Tr(\tilde{w}w) + \beta_3\Tr(wp_4\tilde{p}_4\tilde{w}),\end{array}\ee 

From the explicit form of $M_4$  (\ref{expr4prime}), using (\ref{id2x2})-(\ref{id4x4}), one can write explicitly the first three terms in (\ref{a5potential}) 
\be\label{det2}-\gamma_2(\Qt_1Q_1+\tilde{v}v+\tilde{w}w)+\dots\ee 
\be\label{det3}\gamma_3(Q_1\tilde{w}v+\tilde{Q}_1\tilde{v}w+\tilde{v}p_4\tilde{p}_4v+\tilde{w}p_4\tilde{p}_4w)+\dots\ee 
\be\label{det4}\gamma_4(\tilde{v}v\tilde{w}w-\tilde{w}v\tilde{v}w-\tilde{Q}_1\tilde{v}p_4\tilde{p}_4w-Q_1\tilde{w}p_4\tilde{p}_4v) +\dots\ee 
The dots stand for all terms proportional to the trace of $M_4$, which is just equal to $\tilde{p}_4p_4$. As will become clear shortly, they don't play any role in our analysis, so we do not write them explicitly. This is essentially due to the F-term of $\varphi_5$ (the singlet in the vectormultiplet of the rightmost $U(1)$ node in the lower row in (\ref{step3})), which implies that $p_4\tilde{p}_4$ squares to zero.

The $U(2)$ node now confines and is traded for a $3\times3$ chiral multiplet $Y_3$
\be\label{expr41} Y_3=\left(\begin{array}{ccc}
Y_{11} & Q_2 & Y_{13} \\
\Qt_2 & Y_{22} & Y_{23} \\
Y_{31} & Y_{32} & Y_{33}
        \end{array}\right)
\ee 
which provides one extra $U(1)_{\text{red}}\times U(1)_F$ bifundamental (we named those components $Q_2$ and $\Qt_2$). We have the usual superpotential term $\gamma\det Y_3$ and according to the monopole duality $\M_4^-$ is identified with $\gamma$. The operator $\tilde{p}_4p_4$ is now replaced by $Y_{33}$. At this stage we are left with the theory  

\begin{center}
 \begin{tikzpicture}[->, thick]
\node[shape=circle, draw, minimum height=.8cm] (1) at (0,0) {$1$};
\node[shape=circle, draw, minimum height=.8cm,red,thick] (2) at (-1.3,1.5) {$1$};
\node[rectangle, draw, minimum height=.8cm,thick] (3) at (1.3,1.5) {$\,\,1\,\,$};
\node at (-0.0,1.9) {$Q_{1,2},\!\Qt_{1,2}$};
\draw[-] (1) -- (2);
\draw[-] (1) -- (3);
\draw[-] (-0.9,1.57) -- (0.9,1.57);
\draw[-] (-0.9,1.45) -- (0.9,1.45);
\end{tikzpicture}
\end{center}

\noindent and in terms of the matrix $Y_3$ the superpotential reads 
\bea \nn \label{newsup}\W &= & \gamma_3(Q_1\Qt_2+\Qt_1Q_2+Y_{13}Y_{31}+Y_{23}Y_{32})+\gamma_4(Y_{11}Y_{22}-Q_2\Qt_2-\Qt_1Y_{13}Y_{32}-Q_1Y_{23}Y_{31})\\ 
&&-\gamma_2(\Qt_1Q_1+Y_{11}+Y_{22})+ \varphi_5Y_{33}+\M^+ + \alpha_0\M^- +\beta_2Y_{22}+\beta_3Y_{23}Y_{32}\\ &&+\gamma\det Y_3 + \alpha_1(\gamma + M')+Y_{33}(\dots).\nn \eea
The last term $Y_{33}(\dots)$ denotes all the terms in (\ref{det2})-(\ref{det4}) we did not write explicitly, which are all proportional to $\Tr M_4=Y_{33}$. The monopoles $\M^\pm$ are charged under the topological symmetry of the node with two flavors. The diagonal fields $Y_{ii}$ ($i=1,2,3$) and the singlets $\varphi_5$, $\beta_2$ and $\gamma_2$ are now massive and can be integrated out. The F-term for $\varphi_5$ sets $Y_{33}$ to zero, hence also the last term in the superpotential vanishes:
\bea \W \nn &= & \gamma_3(Q_1\Qt_2+\Qt_1Q_2+Y_{13}Y_{31}+Y_{23}Y_{32})-\gamma_4(Q_2\Qt_2+\Qt_1Y_{13}Y_{32}+Q_1Y_{23}Y_{31})\\ 
&&+\gamma\det Y_3 + \alpha_1(\gamma + M')+\M^+ + \alpha_0\M^- +\beta_3Y_{23}Y_{32} \eea

Finally, when the abelian node with two flavors confines the superpotential term $\gamma_{Y_2}\det Y_2$ is generated, with the $2\times 2$ matrix $Y_2$, whose off-diagonal entries combine into a hypermultiplet charged under the leftover $U(1)_{\text{red}}$ gauge group:
\be\label{expr51} Y_2=\left(\begin{array}{cc}
Y'_{11} & Q_3 \\
\Qt_3 & Y'_{22}     \end{array}\right)
\ee 

 As we mentioned before, the field $M'$ is now identified with $\gamma$ which becomes massive and the terms in the third line of (\ref{newsup}) can be dropped from the superpotential because of F-terms. The superpotential term proportional to $\det Y$ is also set to zero by the F-term for $\alpha_0$ ($\gamma_{Y_2}=0$) and we are left with 
\be\W=\gamma_3(Q_1\Qt_2+\Qt_1Q_2+Y_{11}+Y_{22})-\gamma_4(Q_2\Qt_2+\Qt_1Q_3+Q_1\Qt_3)+\beta_3Y'_{22}.\ee 
The diagonal components of $Y_2$ are massive and can be integrated out. The fields $Q_1,\Qt_1,Q_2,\Qt_2,Q_3$ and $\Qt_3$ survive in the IR, they transform in the bifundamental under the IR $U(1)_{\text{red}}\times U(1)_F$ theory:
\be \nn
\bpic[->, thick]
\node at (-3,0) {$\tilde{\CT}_{IR}:$};
\node[shape=circle, draw, minimum height=.9cm,red,thick] (1) at (-1.5,0) {$1$};
\node[rectangle, draw, minimum height=.9cm,thick] (2) at (1.5,0) {$\,\,1\,\,$};
\draw[-] (1.1,0.15) -- (-1.1,0.15);
\draw[-] (1.1,0) -- (-1.1,0);
\draw[-] (1.1,-0.15) -- (-1.1,-0.15);
\node at (0,0.5) {$Q_{1,2,3},\!\Qt_{1,2,3}$};
\end{tikzpicture}
\qquad \bpic \node at (0,0.3) {\large$=$}; \node at (0,0) {};\epic \qquad
\bpic
\node[shape=circle, draw, minimum height=.9cm,red,thick] (1) at (-1,0) {$1$};
\node[rectangle, draw, minimum height=.9cm,thick] (2) at (1,0) {$\;3\;$};
\draw[-,thick] (1) to (2);
\epic \ee
The superpotential for $\tilde{\CT}_{IR}$
\be\label{finalsup}\W_{\tilde{\CT}_{IR}}=-\gamma_4(Q_2\Qt_2+\Qt_1Q_3+Q_1\Qt_3),\ee  
 is just (modulo a field redefinition) the superpotential of $\CN=4$ SQED with three flavors, as we wanted to show. Notice also that the equations of motion set to zero $\beta_2$ and $\beta_3$, which is consistent with our findings in section \ref{dressed1} that the $\b_j$'s vanish in the chiral ring.

\subsection{Comments about the higher $N$ generalization}
The analysis in the general case proceeds in the same way, although the detailed computation quickly gets involved. In this section we will give the answer for some higher rank cases, namely $A_7$ and $A_9$ AD theories. We will just state the result without providing all the details of the derivation. 

\subsubsection*{$SU(4)$ SQCD and $A_7$ AD theories} 

The UV theory which in 4d flows in the IR to $A_7$ AD and constitutes our starting point is $SU(4)$ adjoint SQCD with eight flavors and superpotential 
\be\W=\sum_{i=1}^8\tilde{q}_i\Phi q^i +\sum_{i=1}^7\tilde{q}_iq^{i+1}+\sum_{i=2}^4\beta_i\Tr\Phi^i+\alpha_0\tilde{q}_8q^1+\alpha_1(\tilde{q}_7q^1+\tilde{q}_8q^2)+\alpha_2(\tilde{q}_6q^1+\tilde{q}_7q^2+\tilde{q}_8q^3).\ee
In the mirror quiver, after the dualization of all the gauge groups of one tail and the central node, we are left with the theory 
\begin{figure}[ht!]\label{su4par}
\begin{center}
 \begin{tikzpicture}[->, thick]
\node[shape=circle, draw, minimum height=.9cm] (9) at (-2,0) {$1$}; 
\node[shape=circle, draw, minimum height=.9cm] (1) at (0,0) {$2$};
\node[shape=circle, draw, minimum height=.9cm] (2) at (2,0) {$3$};
\node[shape=circle, draw, minimum height=.9cm,red,thick] (3) at (4,1) {$1$};
\node[shape=rectangle, draw, minimum height=.7cm] (4) at (4,-1) {$\;1\;$};
\node[] (5) at (1,0.3) {$p,\tilde{p}$};
\node[] (6) at (3,0.8) {$v,\tilde{v}$}; 
\node[] (7) at (3,-0.9) {$w,\tilde{w}$}; 
\node[] (8) at (4.65,0) {$Q_1,\tilde{Q}_1$};

\draw[-] (1) -- (2);
\draw[-] (2) -- (3);
\draw[-] (2) -- (4);
\draw[-] (4) -- (3);
\draw[-] (1) -- (9);
; 
\end{tikzpicture}
\end{center}
\end{figure}

In this quiver the operator $p\tilde{p}$, which is a $U(3)$ adjoint, satisfies the chiral ring relation $(p\tilde{p})^3=0$. This is due to the F-term relations of the linear tail. When the gauge group $U(4)$ confines we are left with a $5\times 5$ chiral $M_5$, which in terms of the fields appearing in (\ref{su4par}) takes the form 
\be\label{expr5} M_5=\left(\begin{array}{ccc}
0 & Q_1 & \tilde{v} \\
\tilde{Q}_1 & 0 & \tilde{w} \\
v & w & p\tilde{p}
        \end{array}\right)
\ee 
Along the way we generate the superpotential terms $\gamma_i\det M_i$, whose form can be derived using (\ref{iddet}). The terms proportional to $\gamma_2$ and $\gamma_3$ are exactly as in (\ref{det2}), (\ref{det3}) (with $p_4\tilde{p}_4$ replaced by $p\tilde{p}$) so we don't write them again. The term involving $\gamma_4$ is as in (\ref{det4}) except for two extra terms: 
\be\label{ddet4}\gamma_4(\tilde{v}v\tilde{w}w-\tilde{w}v\tilde{v}w-\tilde{Q}_1\tilde{v}p\tilde{p}w-Q_1\tilde{w}p\tilde{p}v-\tilde{w}(p\tilde{p})^2w-\tilde{v}(p\tilde{p})^2v).\ee 
The term involving $\gamma_5$, namely the determinant of $M_5$, reads 
\be\label{det5}\gamma_5(\tilde{Q}_1\tilde{v}(p\tilde{p})^2w+Q_1\tilde{w}(p\tilde{p})^2v+\tilde{w}v\tilde{v}p\tilde{p}w+\tilde{v}w\tilde{w}p\tilde{p}v-\tilde{v}v\tilde{w}p\tilde{p}w-\tilde{w}w\tilde{v}p\tilde{p}v).\ee 
Dualizing the remaining three gauge nodes with monopole superpotential we generate the surviving fields $Q_2,Q_3,Q_4$ and land on the theory 
\begin{center}
 \begin{tikzpicture}[->, thick]
\node[shape=circle, draw, minimum height=.9cm,red,thick] (1) at (-1,0) {$1$};
\node[shape=rectangle, draw, minimum height=.9cm] (2) at (1,0) {$\;1\;$};
\node[] (5) at (2.5,0) {\Large$=$};
\node[shape=circle, draw, minimum height=.9cm,red,thick] (3) at (4,0) {$1$};
\node[rectangle, draw, minimum height=.9cm,thick] (4) at (6,0) {$\;4\;$};
\draw[-,thick] (3) to (4);
\draw[-] (0.65,0.3) -- (-0.7,0.3);
\draw[-] (0.65,-0.3) -- (-0.7,-0.3);
\draw[-] (0.65,0.1) -- (-0.55,0.1);
\draw[-] (0.65,-0.1) -- (-0.55,-0.1);
; 
\end{tikzpicture}
\end{center}

namely SQED with four flavors and superpotential 
\be\W=\gamma_5(\tilde{Q}_1 Q_4+\tilde{Q}_2 Q_3+\tilde{Q}_3 Q_4+ \tilde{Q}_4 Q_1).\ee 

\subsubsection*{$SU(5)$ SQCD and $A_9$ theory} 

In order to engineer $A_9$ AD theory we start from $\CN=4$ $SU(5)$ SQCD with 10 flavors and modify the superpotential according to the procedure discussed above. After the dualization of all the gauge nodes in one tail and the central node, we find the model 
\begin{figure}[ht!]\label{su5par}
\begin{center}
 \begin{tikzpicture}[->, thick]
 \node[shape=circle, draw, minimum height=.9cm] (10) at (-4,0) {$1$}; 
\node[shape=circle, draw, minimum height=.9cm] (9) at (-2,0) {$2$}; 
\node[shape=circle, draw, minimum height=.9cm] (1) at (0,0) {$3$};
\node[shape=circle, draw, minimum height=.9cm] (2) at (2,0) {$4$};
\node[shape=circle, draw, minimum height=.9cm,red,thick] (3) at (4,1) {$1$};
\node[shape=rectangle, draw, minimum height=.7cm] (4) at (4,-1) {$\;1\;$};
\node[] (5) at (1,0.3) {$p,\tilde{p}$};
\node[] (6) at (3,0.8) {$v,\tilde{v}$}; 
\node[] (7) at (3,-0.9) {$w,\tilde{w}$}; 
\node[] (8) at (4.6,0) {$\,Q_1,\tilde{Q}_1$};

\draw[-] (1) -- (2);
\draw[-] (2) -- (3);
\draw[-] (2) -- (4);
\draw[-] (4) -- (3);
\draw[-] (1) -- (9);
\draw[-] (9) -- (10);
; 
\end{tikzpicture}
\end{center}
\end{figure}

Again the central node is confined and can be traded for a $6\times 6$ chiral multiplet $M_6$ which reads 
\be\label{expr6} M_6=\left(\begin{array}{ccc}
0 & Q_1 & \tilde{v} \\
\tilde{Q}_1 & 0 & \tilde{w} \\
v & w & p\tilde{p}
        \end{array}\right)
\ee 
The $U(4)$ adjoint $p\tilde{p}$ now satisfies the constraint $(p\tilde{p})^4=0$. In terms of these fields the superpotential terms proportional to $\gamma_2$ and $\gamma_3$ are as in (\ref{det2}) and (\ref{det3}) respectively. The term proportional to $\gamma_4$ is as in (\ref{ddet4}) and the one proportional to $\gamma_5$ is 
\be\gamma_5(\tilde{v}(p\tilde{p})^3v+\tilde{w}(p\tilde{p})^3w+\dots),\ee 
where the dots stand for all the terms appearing in (\ref{det5}). The determinant of $M_6$ reads 
\be\begin{array}{ll} &\tilde{w}w\tilde{v}(p\tilde{p})^2v+\tilde{v}v\tilde{w}(p\tilde{p})^2w-\tilde{v}w\tilde{w}(p\tilde{p})^2v- \tilde{w}v\tilde{v}(p\tilde{p})^2w+(\tilde{v}p\tilde{p}v)(\tilde{w}p\tilde{p}w)-\\ &(\tilde{w}p\tilde{p}v)(\tilde{v}p\tilde{p}w)-Q_1\tilde{w}(p\tilde{p})^3v-\tilde{Q}_1\tilde{v}(p\tilde{p})^3w.\end{array}\ee
When the gauge nodes in the linear tail in (\ref{su5par}) confine, only the superpotential term proportional to $\gamma_6$ survives and we are left with SQED with five flavors $Q_i, \Qt_i$ $i=1,2,\ldots,5$. In terms of these fields the superpotential reads 
\be\W=-\gamma_6\left(\tilde{Q}_1 Q_5 + \tilde{Q}_2 Q_4+ \tilde{Q}_3 Q_3 +\tilde{Q}_4 Q_2+\tilde{Q}_5 Q_1 \right).\ee 

\subsection{The Maruyoshi-Song procedure in $3d$} \label{MS3d}

If we repeat the procedure of Maruyoshi and Song in $3d$, we find that more $\a_r$ fields remain coupled to the $SU(N)$ gauge theory. In order to identify them, we can use algebraic arguments instead of performing $\CZ$-extremizations. Moreover, the conclusions are valid both in $3d$ and $4d$.

Let us start from the $8$-supercharges theory $SU(N)$ with $2N$ flavors and couple a $2N \times 2N$ matrix to the Higgs Branch moment map. After giving a maximal nilpotent vev to $A$, the superpotential is given by eq. (\ref{IRWnilpotent}). It contains many terms, not necessarily linear in $\a_r$. As we explain in \ref{nilpotvev}, chiral ring stability \cite{Benvenuti:2017lle} arguments, analogous to those given in section \ref{ABELIAN}, imply that all the terms containing $\Tr(\qt \phi^r q)$ drop out in the IR if $r\geq N$. The remaining superpotential contains $N-1$ $\a_r$'s (the others decouple) and is simply
\be\label{cciao}\W=\sum_{r=0}^{N-1}\alpha_r \Tr(\qt\phi^rq).\ee 
Performing $a$-maximization in $4d$, it turns out that $\a_{N-1}$ decouples from the theory.

We can also think of a theory with the superpotential \ref{cciao} as the naive compactification to $3d$ of the $4d$ theory: we start in $4d$ from $SU(N)$ with $2N$ flavor coupled to $A$ which takes a nilpotent vev, and compactify to $3d$ before flowing to the IR. The arguments given in section \ref{compact} imply that also in this case no monopole superpotential terms are generated in the compactification.

In $3d$, the difference between \ref{cciao} and the theory $\CT'_{3d,IR}$, analyzed in detail in section \ref{ABELIAN}, is the presence of the superpotential term $\alpha_{N-1}\Tr(\qt\phi^{N-1}q)$. The crucial point is that the singlet $\a_{N-1}$ does not decouple from the rest of the theory. In analogy with \cite{Benvenuti:2017lle}, the $3d$ theory with superpotential (\ref{cciao}) abelianizes to $U(1)^{N-1}$ linear quiver and $N$ singlet fields that flip each meson. Instead of (\ref{N=4quiver}),  $\CT_{3d,IR}$ is dual to the $\CN=2$ quiver
\be \label{N=2quiver}\bpic \path (-4,-3.2) node[rectangle,draw](z1) {\,1\,} -- (-3,-3.2) node[circle,draw](z2) {\!1\!} -- (-2,-3.2) node(z3) {$\cdots$} -- (-1,-3.2) node[circle,draw](z4) {\!1\!} -- (-0,-3.2) node[rectangle,draw](z5) {\,1\,} -- (2,-3.2) node[right] {$\CW\!=\sum_{i=1}^{N} \Phi_i P_i\Pt_i $}; 
\draw [-] (z1) to (z2);
\draw [-] (z2) to (z3);
\draw [-] (z3) to (z4);
\draw [-] (z4) to (z5);
\epic\ee
Here the 'long mesons' $\prod P_i$ and $\prod \Pt_i$ have vanishing product, due to the $\CF$-terms of $\Phi_i$'s. This is consistent with the properties of the $SU(N)$ model, in which the $\CF$-terms of $\a_{N-1}$ set to zero $\Tr(\qt\phi^{N-1}q)$, the operator we called $\CM$ in section \ref{dressed1}. As a result, the chiral ring relation (\ref{ccad}) $\CB\tilde{\CB}=\CM$ in $\CT'_{3d,IR}$ becomes $\CB\tilde{\CB}=0$ in $\CT_{3d,IR}$.

We can at this point define a $\CT_{3d,UV}$ involving the fields $\a_r$ $r=0,\ldots,N-1$, analogous to $\CT'_{3d,UV}$: this is a $SU(N)$ theory with $2N$ flavors, an adjoint $\phi$ and superpotential 
\be\W=\sum_{i=1}^{2N} \Tr(\qt_i \phi q^i) + \sum_{i=1}^{2N-1} \Tr(\qt_i q^{i+1})+\sum_{r=0}^{N-1}\sum_{i=0}^r \a_r \Tr(\qt_{2N+i-r}q^{i+1}),\ee 
which reduces precisely to (\ref{cciao}) upon integrating out massive flavors (see also Appendix \ref{nilpotvev}).
We will now discuss the mirror dual of the RG flow $\CT_{3d,UV} \rightarrow \CT_{3d,IR}$. 

In the case $N=3$, by repeating the analysis of section \ref{su3mm} for $\CT_{3d,UV}$, we find that the mirror theory still reduces to SQED with 3 flavors: the mirror of $\CT_{3d,UV}$ has superpotential (using the same notation as in section \ref{su3mm})
\be\label{mirrorn}\begin{array}{ll}\W=&\W_{\CN=4}+\sum_{i=1}^5 \M_i^+ + \alpha_0\M^{-----} + \alpha_1(\M^{----0}+\M^{0----})  + \\ 
&\alpha_2(\M^{---00}+\M^{0---0}+\M^{00---}).\end{array}\ee 
We should now repeat the procedure explained in section \ref{su3mm}, replacing (\ref{mirror3}) with the above equation. All the gauge groups in the lower row of (\ref{sqcd3}) confine as before and the monopole operators appearing in the second row of (\ref{mirrorn}) are identified with $\gamma_4$. As a result (\ref{finalsup}) is replaced by 
\be\W=-\gamma_4(Q_2\Qt_2+\Qt_1Q_3+Q_1\Qt_3)+3\a_2\gamma_4.\ee
Since the $\beta_{2,3}$ terms in this case are absent, the singlets $Y_{22}$ and $Y'_{22}$ appearing in section \ref{su3mm} decouple and become free instead of acquiring mass. 

The crucial difference with respect to the analysis of section \ref{su3mm} is that $\alpha_2$ makes the singlet $\gamma_4$ massive and the superpotential vanishes. All other singlets $\alpha_r$ and $\gamma_i$ still become massive. The same conclusion holds for arbitrary $N$: the singlet $\alpha_{N-1}$ makes $\gamma_{N+1}$ massive and we are left with $\CN=2$ SQED with $N$ flavors and no superpotential. In conclusion, we find that $\CT_{3d,UV}$ flows in the IR to the mirror of $\CN=2$ SQED, which is precisely the abelian linear quiver discussed around (\ref{N=2quiver}) \cite{Aharony:1997bx}.   

\acknowledgments{We are grateful to Francesco Benini, Matthew Buican, Amihay Hanany and Alberto Zaffaroni for useful discussions and comments. S.B. is partly supported by the INFN Research Projects GAST and ST$\&$FI and by PRIN 'Geometria delle varieta algebriche'. The research of S.G. is partly supported by the INFN Research Project ST\&FI.}

\appendix

\section{Nilpotent vevs}\label{nilpotvev}

In this section we discuss, following \cite{Agarwal:2014rua}, the superpotential generated by the Maruyoshi-Song procedure for $SU(N)$ with $2N$ flavors. When we turn on a nilpotent vev for the matrix of flipping fields $A$, in the form of a single Jordan block of size $2N$, we break the $SU(2N)$ symmetry completely, leaving just the baryon number unbroken. In the resulting RG flow some chiral multiplets decouple and in the IR we are left with a free sector consisting of decoupled chiral multiplets plus an 
interacting theory which turns out to be equivalent to $(A_1,A_{2N-1})$. 

As is well-known, $SU(N)$ nilpotent orbits are in one-to-one correspondence with $SU(2)$ embeddings $\rho$ into $SU(N)$. For every such 
embedding $\rho(\sigma^+)$ is nilpotent and we can assume it is in Jordan form, with blocks of size $n_i$. 
Under the above mentioned embedding, the fundamental representation of $SU(N)$ decomposes into irreducible 
representations of $SU(2)$ as $\mathbf{N}\rightarrow 
\sum_{i=1}^l\mathbf{n}_i$. We can easily derive from this formula the decomposition 
of the adjoint of $SU(N)$:
\be\label{decom} \mathbf{adj.}=\bigoplus_{i=1}^{l}\bigoplus_{s=1}^{n_i-1}V_{s}\oplus(l-1)V_0\oplus2\left[ 
\bigoplus_{i<j}\bigoplus_{k=1}^{n_j}V_{\frac{n_i+n_j-2k}{2}}\right]\ee 
where $V_{s}$ is the spin $s$ representation of $SU(2)$. 
When we turn on a nilpotent vev of the form $$\langle A\rangle=\rho(\sigma^+),$$ 
we break spontaneously the global symmetry down to the commutant of $SU(2)$ inside $SU(N)$. By expanding the superpotential around 
the vev we find 
\be\label{pot}\mathcal{W}=\Tr\rho(\sigma^+)\mu+\Tr A\mu.\ee 
The first term is the source of global symmetry breaking and as a result several components $\mu_i$ of the $SU(N)$ moment map (in our case the meson) 
will combine with the current multiplets into long multiplets. The components of the flipping field $A$ coupled to $\mu_i$'s will 
now decouple and become free. These are the Goldstone multiplets associated with the spontaneous symmetry breaking. 

How can we determine which components $\mu_i$ decouple? Given the form of the superpotential, we can observe that under an infinitesimal 
complexified $SU(N)$ transformation we can obtain all the components of $\mu$ except those which commute with $\rho(\sigma^+)$. 
On the other hand, since $\rho(\sigma^+)$ is the $SU(2)$ raising operator, we immediately conclude that the only components of the moment 
map which commute with it are the highest weight states in each $V_s$ appearing in (\ref{decom}). Accordingly, the only components of 
$A$ which remain coupled to the theory are the lowest states of each $SU(2)$ irreducible representation. 

In writing the superpotential as in (\ref{pot}), we should keep only the components of the flipping field which do not decouple. 
In the case relevant for AD theories, a single Jordan block of size $2N$, $A$ (actually its vev plus fluctuations around it) takes the form:
\begin{equation}\label{nvev}
A=\left(\begin{array}{ccccc}
  0 & 1 & 0 & \dots & 0 \\
  \a_{2N-2} & 0 & 1 & \dots & 0\\
  \a_{2N-3} & \a_{2N-2} & \ddots & \ddots & 0\\
  \vdots & \ddots & \ddots & 0 & 1\\
  \a_0 & \dots & \a_{2N-3} & \a_{2N-2} & 0
\end{array}\right)
 \end{equation} 
The vev of the flipping field indeed breaks the UV R-symmetry, which is now mixed with $\rho(\sigma_3)$. After the vev, the trial 
R-symmetry should then be redefined by subtracting $(1+\epsilon)\rho(\sigma_3)$. The value of $\epsilon$ can be found performing 
a-maximization. 
 
In order to complete the analysis, we take into account the fact that the vev for $A$ gives mass to all the $SU(N)$ fundamentals 
except one and we should integrate out all the massive multiplets. This can be done following the procedure described in \cite{Agarwal:2014rua}: the superpotential becomes 
\be \label{IRWnilpotent}\mathcal{W}=Z\widetilde{Z}\Phi+A\widetilde{Z}Z+\sum_{n=1}^{2N-1}(ZAB^n\widetilde{Z}+ZB^n\widetilde{Z}\Phi), \ee 
where $A$ is as in (\ref{nvev}), $\Phi$ is the $SU(N)$ adjoint, $Z$ and $\widetilde{Z}$ are the massless fields, in our case 
$$\widetilde{Z}=\left(\begin{array}{c}
                         \widetilde{Q}_1 \\ 
                         0 \\
                         \vdots \\
                         0
                        \end{array}\right);\quad Z=(0\dots,0,Q_{2N}),$$ 
where we suppressed the color indices, and the matrix $B$ is 
\be B=-\left(\begin{array}{ccccc}
 0 & & & \dots & 0\\
 \Phi& 0 & & \dots & 0\\
 \a_{2N-2} & \Phi & 0 & \dots & 0\\
 \vdots & \ddots & \ddots & \ddots & \vdots\\
 \a_{1} & \dots & \a_{2N-2} & \Phi & 0
   \end{array}\right).\ee 
More explicitly, the cubic and quartic terms have the following form: 
\be\label{massiv}\mathcal{W}=Q_{2N} \a_0\widetilde{Q}_1-Q_{2N}(2\a_{1}\Phi+\sum_{k=2}^{2N-2}\a_{2N-k}\a_{k})\widetilde{Q}_1+\dots\ee
At the cubic level only the singlet $\a_0$ appears. The other $2N-2$ chiral multiplets $\a_{r}$ ($r=1,\dots,2N-2$) appear only 
in quartic or higher terms. 

Using the chiral ring stability criterion of \cite{Benvenuti:2017lle}, we can significantly simplify (\ref{massiv}): first of all we notice that the $N \times N$ matrix $\phi$ satisfies the charachteristic polynomial equation, so $\widetilde{Q}_1\Phi^{j \geq N}Q_{2N}$ can be written as a polynomial in $\widetilde{Q}_1\Phi^{j < N}Q_{2N}$ and the Casimirs of $\Phi$. Then we can notice that $\alpha_0$ and $\a_1$ appear only in the terms $\a_0\widetilde{Q}_1Q_{2N}$ and $\a_1\widetilde{Q}_1\Phi Q_{2N}$ respectively, so their F-terms set to zero $\widetilde{Q}_1Q_{2N}$ and $\widetilde{Q}_1\Phi Q_{2N}$. This implies that all other terms of the form $\widetilde{Q}_1Q_{2N}(\dots)$ and $\widetilde{Q}_1\Phi Q_{2N}(\dots)$ such as the last term in (\ref{massiv}) can be dropped. At this stage it is straightforward to check that the only surviving term containing $\a_2$ is $\a_2\widetilde{Q}_1\Phi^2Q_{2N}$. Combining the F-terms for $\a_0$ and $\a_2$, which reads 
\be\widetilde{Q}_1\Phi^2Q_{2N}=\a_{2N-2}\widetilde{Q}_1Q_{2N},\ee 
we conclude that $\widetilde{Q}_1\Phi^2Q_{2N}$ is zero in the chiral ring, hence all terms proportional to this operator can be dropped. Proceeding recursively in this way, we find that the F-terms for $\a_r$ with $r<N$ set to zero all dressed mesons of the form $\widetilde{Q}_1\Phi^{j < N}Q_{2N}$. Consequently, operators of the form $\widetilde{Q}_1\Phi^{j \geq N}Q_{2N}$ automatically vanish in the chiral ring because of the characteristic polynomial constraint. The conclusion is that the term  $\widetilde{Q}_1\Phi^{2N}Q_{2N}$ can be removed and all the singlets $\a_r$ with $r\geq N$ disappear from the superpotential and decouple. This observation tells us that the superpotential reduces to the simpler form
\be\W=\sum_{r=0}^{N-1}\a_r\widetilde{Q}_1\Phi^{r}Q_{2N}.\ee 
This argument is valid in any spacetime dimension. Of course the set of operators which violate the unitarity bound and decouple is dimension dependent: in 4d $\a_{N-1}$ decouples and the corresponding superpotential term drops out, whereas in 3d all the singlets are above the unitarity bound.



\bibliographystyle{ytphys}

\end{document}